\documentstyle[11pt]{article}
\newcommand{\hm}{\hspace{-0.1in}}

\newcommand{\beq}{\begin{equation}}
\newcommand{\beqar}{\begin{eqnarray}}
\newcommand{\eeq}[1]{\label{#1} \end{equation}}
\newcommand{\eeqar}[1]{\label{#1} \end{eqnarray}}

\begin{document}
\renewcommand{\topfraction}{0.95}
\renewcommand{\textfraction}{0.05}
\pagestyle{myheadings}
\thispagestyle{empty}
\def\ti#1 {\begin{center} \baselineskip=17pt {\large #1} \end{center}}
\markboth
{\it P. M\"{o}ller, J. R. Nix, W. D. Myers, and W. J. Swiatecki/Nuclear Masses}
{\it P. M\"{o}ller, J. R. Nix, W. D. Myers, and W. J. Swiatecki/Nuclear Masses}
\mbox{ } \vspace{1.0in} \mbox{ }\\
\ti{In Memoriam}
\hspace*{\parindent}This paper is dedicated to the memory of our friend and
colleague John L.  Norton, who wrote the original versions of the computer
programs that we use to calculate the single-particle energies and resulting
shell and pairing corrections for a deformed folded-Yukawa single-particle
potential.
\newpage
\pagestyle{myheadings}
\thispagestyle{empty}
\mbox { } \\
\newpage
\thispagestyle{empty}
\setcounter{page}{1}
\mbox{ } \hfill \mbox{ } \\
\mbox{ } \vspace{-0.2in} \mbox{ } \\
\begin{center}
\begin{large}
{\bf NUCLEAR GROUND-STATE MASSES AND DEFORMATIONS}
\\[2ex]\end{large}
P. M\"{o}ller and J. R. Nix\\
Theoretical Division, Los Alamos National
Laboratory, Los Alamos, NM 87545\\[2ex]
W. D. Myers and W. J. Swiatecki\\
Nuclear Science Division, Lawrence Berkeley
Laboratory, Berkeley, CA 94720
   \\[2ex]
August 16, 1993 \\[3ex]
\end{center}
\begin{small}
\begin{description}
\item[Abstract:]
We tabulate the atomic mass excesses and nuclear ground-state deformations of
8979 nuclei ranging from $^{16}$O to $A=339$.  The calculations are based on
the finite-range droplet macroscopic model and the folded-Yukawa
single-particle microscopic model. Relative to our 1981 mass table the current
results are obtained with an improved macroscopic model, an improved pairing
model with a new form for the effective-interaction pairing gap, and
minimization of the ground-state energy with respect to additional shape
degrees of freedom.  The values of only 9 constants are determined directly
from a least-squares adjustment to the ground-state masses of 1654 nuclei
ranging from $^{16}$O to $^{263}$106 and to 28 fission-barrier heights.  The
error of the mass model is 0.669~MeV for the entire region of nuclei
considered, but is only 0.448~MeV for the region above $N=65$.\\[1ex]

\end{description}
\end{small}
\section{Introduction}

We presented our
first macroscopic-microscopic global
nuclear mass calculation 12 years ago$\,^{1,2})$.
This calculation, which was based on a finite-range liquid-drop model
for the macroscopic energy and a folded-Yukawa single-particle
potential for the microscopic corrections,
was somewhat limited in scope.
With only 4023 nuclei included, it did not extend to the proton or
neutron drip lines or to the region of superheavy nuclei.
Also, the quantities tabulated were
limited to ground-state masses, $Q_2$ and $Q_4$ moments, and
microscopic corrections.

Our next publication of calculated nuclear masses occurred five years
ago$\,^{3,4})$.  In these calculations new pairing
models had been incorporated and two different macroscopic models were
investigated, namely the finite-range liquid-drop model
(FRLDM)$\,^{3})$ and the finite-range droplet model
(FRDM)$\,^{4})$.  These abbreviations are also used to designate
the full macroscopic-microscopic nuclear structure models based on the
respective macroscopic models.  The former is the macroscopic model used
in the 1981$\,^{1,2})$ calculations and the latter is
an improved version$\,^{5})$ of the droplet
model$\,^{6-8})$.  Because
there were several unresolved issues in the 1988
calculations$\,^{3,4})$ these tables should be regarded
as interim progress reports.

We have now resolved these issues, which were related to the pairing
calculations$\,^{9})$, to the effect of higher-multipole
distortions on the ground-state mass$\,^{10})$, and to some
details of the shell-correction and zero-point-energy
calculations$\,^{11})$.  The resolution of these issues has
resulted in the present mass table.  We first briefly review some
important results obtained in the original 1981 calculation and
enumerate the additional features of our new calculations.

Subsequent comparisons of predictions of our original 1981
model$\,^{1,2})$ with nuclear masses measured after the
calculations were published showed that the model would reliably predict
masses of nuclei that were not included in the determination of model
constants$\,^{3,11})$.  With a properly defined model
error$\,^{3})$, the error in new regions of nuclei is about the
same as in the region where the constants were adjusted.  In the most
recent investigations$\,^{11,12})$
of the 1981 mass calculation in new regions of
nuclei, the error for 351 new nuclei was only 6\% larger than the error
in the region where the model constants were adjusted.
Furthermore,
the error did not increase with distance from $\beta$ stability.

Also, many other nuclear-structure properties were successfully
predicted by the model for nuclei far from
stability$\,^{13-16})$.  A
special result of the 1981 mass calculation was the interpretation of
certain spectroscopic results in terms of an intrinsic octupole
deformation of nuclei in their ground
state$\,^{1,17-19})$.

Here we present results of our new calculations of nuclear ground-state
masses and deformations.  Relative to the 1981 calculations we use an
improved macroscopic-microscopic model, include additional shape degrees
of freedom, extend the calculations to new regions of nuclei, and
calculate a large number of additional nuclear ground-state properties.
These additional properties will be published in a forthcoming article
devoted to nuclear astrophysics$\,^{20})$.

Specifically, we have improved the model in the following areas:
\begin{itemize}

\item
Our preferred macroscopic model is now the finite-range droplet model,
which contains several essential
im\-prove\-ments$\,^{4,5})$ relative to the original
droplet model$\,^{6-8})$.

\item
The pairing calculations have been improved. Our pairing model
is now the Lipkin-Nogami model$\,^{21-23})$.
We also use an improved functional form of the effective-interaction
pairing gap and an optimized pairing
constant$\,^{9,24})$.

\item An eighth-order Strutinsky shell correction is used.

\item The $\epsilon$ zero-point energy is still added
to the calculated potential energy
to obtain the ground-state mass, but no $\gamma$ zero-point
energy is added, since the method of calculation is not
sufficiently accurate$\,^{11})$.

\item
We minimize the ground-state energy with respect to $\epsilon_3$
and $\epsilon_6$ shape degrees of freedom, in addition to the
$\epsilon_2$  and $\epsilon_4$ shape degrees of freedom considered
previously.

\item
Each ground-state shell-plus-pairing correction is based on
single-particle levels calculated for the constants appropriate
to the nucleus studied. Earlier, a single set of single-particle levels
was used for an extended region of nuclei
in conjunction  with an interpolation
scheme to improve accuracy.

\item
The calculation has been extended from 4023 nuclei
to 8979 nuclei, which now includes
nuclei between the proton and  neutron drip lines and superheavy
nuclei up to $A=339$.
\end{itemize}.

In the macroscopic-microscopic approach it is possible to calculate
a large number of nuclear structure properties
in addition to  nuclear ground-state masses.
These include the following:
\begin{description}
\item[]
{\bf Even-multipole ground-state deformations:}\\
Quadrupole $\epsilon$ deformation        \hfill $\epsilon_2$ \\
Hexadecapole $\epsilon$ deformation      \hfill $\epsilon_4$ \\
Hexacontatetrapole $\epsilon$  deformation  \hfill  $\epsilon_6$ \\
Related quadrupole $\beta$ deformation        \hfill $\beta_2$ \\
Related hexadecapole $\beta$ deformation      \hfill $\beta_4$ \\
Related hexacontatetrapole $\beta$ deformation  \hfill $\beta_6$
\item[]
{\bf Beta-decay properties:}\\
$Q$ value of the $\beta$ decay        \hfill $Q_{\beta}$  \\
$\beta$-decay half-life               \hfill $T_{1/2}^{\beta}$\\
$\beta$-delayed one-neutron emission probability     \hfill $P_{\rm 1n}$ \\
$\beta$-delayed two-neutron emission probability     \hfill $P_{\rm 2n}$ \\
$\beta$-delayed three-neutron emission probability   \hfill $P_{\rm 3n}$ \\
\item[]
{\bf Lipkin-Nogami pairing quantities:}\\
Neutron pairing gap                   \hfill $\Delta_{\rm n}$   \\
Proton pairing gap                    \hfill $\Delta_{\rm p}$   \\
Neutron number-fluctuation constant   \hfill $\lambda_{\rm 2n}$ \\
Proton number-fluctuation constant    \hfill $\lambda_{\rm 2p}$ \\
\item[]
{\bf Odd-particle spins:}\\
Projection of the odd-neutron angular momentum along the
symmetry axis                              \hfill $\Omega_{\rm n}$   \\
Projection of the odd-proton angular momentum along the
symmetry axis                         \hfill $\Omega_{\rm p}$   \\
\item[]
{\bf Alpha-decay properties:}\\
$Q$ value of the $\alpha$ decay       \hfill $Q_{\rm \alpha}$   \\
$\alpha$-decay half-life                \hfill $T_{1/2}^{\alpha}$ \\
\item[]
{\bf Octupole properties:}\\
Octupole $\epsilon$ deformation           \hfill $\epsilon_3$       \\
Related octupole $\beta$   deformation            \hfill $\beta_3$          \\
Decrease in mass due to octupole
deformation                               \hfill $\Delta E_{3}$\\
\item
{\bf FRDM mass-related quantities:} \\
Spherical macroscopic energy          \hfill    $E_{\rm mac}^{\rm sph}$ \\
Shell correction                      \hfill    $E_{\rm shell}$         \\
Pairing correction                        \hfill    $E_{\rm pair}$           \\
Microscopic correction                \hfill    $E_{\rm mic}$           \\
Calculated mass excess                \hfill    $M_{\rm th}$             \\
Experimental mass excess              \hfill    $M_{\rm exp}$            \\
Experimental uncertainty              \hfill    $\sigma_{\rm exp}$       \\
Discrepancy                           \hfill    $\Delta M$               \\
Calculated binding energy             \hfill    $B_{\rm th}$             \\
\item[]
{\bf FRLDM mass-related quantities:} \\
Finite-range liquid-drop model microscopic
        correction                        \hfill    $E_{\rm mic}^{\rm
FL}$\\[0.02in]
Finite-range liquid-drop model mass excess \hfill   $M_{\rm th}^{\rm FL}$\\
\newpage
\item[]
{\bf Neutron and proton separation energies:} \\
One-neutron separation energy              \hfill   $S_{\rm 1n}$ \\
Two-neutron separation energy              \hfill   $S_{\rm 2n}$ \\
Three-neutron separation energy            \hfill   $S_{\rm 3n}$ \\
One-proton separation energy               \hfill   $S_{\rm 1p}$ \\
Two-proton separation energy               \hfill   $S_{\rm 2p}$ \\

\end{description}
As mentioned above, we present here the calculated ground-state masses and
deformations.
Some of the remaining quantities will be presented in a
forthcoming publication$\,^{20})$.

In the next section we specify the
macroscopic-microscopic finite-range droplet model in some
detail. We discuss in particular the constants of the model,
paying special attention to how to count the number of constants
of a model. We present a summary of {\it all} constants in the model,
including
both those constants that have been determined from a least-squares adjustment
to ground-state masses and fission-barrier heights and those that have been
determined from other considerations.
After our model has been specified, we discuss how it has been applied
to the
current mass calculation.

\section{Models}

In the macroscopic-microscopic method the
total potential energy, which is calculated as
a function of shape, proton number $Z$\/, and neutron number $N$\/, is the
sum of a macroscopic term and a microscopic term
representing the shell-plus-pairing correction.
Thus, the total nuclear
potential energy can be written as
\beq
E_{\rm pot}(Z,N,{\rm shape})= E_{\rm mac}(Z,N,{\rm shape})+
E_{\rm s+p}(Z,N,{\rm shape})
\eeq{toten}
We study here two alternative models for $ E_{\rm mac}$,
given by Eqs.~(\ref{macener}) and (\ref{macenera}) below.
The shell-plus-pairing correction is given
by Eqs.~(\ref{sumspp}) and (\ref{emicr}) below.

It is practical to define an additional energy, the microscopic
correction $E_{\rm mic}$, which is different from the shell-plus-pairing
correction $E_{\rm s+p}$.  For a specific deformation $\epsilon_{\rm
a}$, the latter is determined solely from the single-particle level
spectrum at this deformation by use of Strutinsky's shell-correction
method$\,^{25,26})$ and a pairing model.  In
contrast, the microscopic correction is given by
\beq
E_{\rm mic}( \epsilon_{\rm a}) =
 E_{\rm s+p}( \epsilon_{\rm a}) +
E_{\rm mac}( \epsilon_{\rm a}) -
E_{\rm mac}( \epsilon_{\rm sphere})
\eeq{defmic}
This definition has the desirable consequence that the
potential energy $E_{\rm pot}$ of
a nucleus at a certain deformation, for example the ground-state
deformation $\epsilon_{\rm gs}$,
is simply
\beq
E_{\rm pot}( \epsilon_{\rm gs})  =
E_{\rm mic}( \epsilon_{\rm gs}) +
E_{\rm mac}( \epsilon_{\rm sphere})
\eeq{defpot}
However, the reader should note that the term microscopic correction is
sometimes used instead for shell-plus-pairing correction.
When results are presented it is usually $E_{\rm mic}$ that is tabulated,
because it represents all additional effects over and above the
{\it spherical} macroscopic energy. In practical calculations it is
$E_{\rm s+p}$ that is calculated. To obtain the total energy
a {\it deformed} macroscopic energy term  is then added to
$E_{\rm s+p}$.

There exist several different models for both the macroscopic and
microscopic terms.  Most initial work following the advent of
Strutinsky's shell correction method used the {\it liquid-drop
model}$\,^{27,28})$ as the macroscopic model.

The preferred model in the current calculations has its origin in a 1981
nuclear mass model$\,^{1,2})$, which utilized the
folded-Yukawa single-particle potential developed in
1972$\,^{29,30})$.  The macroscopic model used in the
1981 calculation was a finite-range liquid-drop model, which contained a
modified surface-energy term to account for the finite range of the
nuclear force.  The modified surface-energy term was given by the
Yukawa-plus-exponential finite-range model$\,^{31})$.  The
macroscopic part in this formulation does not describe such features as
nuclear compressibility and corresponding variations in the proton and
neutron radii.

The droplet model$\,^{6-8})$, an extension of
the  liquid-drop model$\,^{27,28})$ that includes
higher-order terms in $A^{-1/3}$ and $(N-Z)/A$, does describe such
features.  However, in its original formulation the droplet model was
very inaccurate for nuclei far from stability and also failed
catastrophically$\,^{31})$ to reproduce fission barriers of
medium-mass nuclei.  These deficiencies led Myers to suggest that the
surface-energy terms of the droplet model also be generalized to account
for the finite range of the nuclear force.  Thus, the
Yukawa-plus-exponential model for the surface tension was incorporated
into the droplet model.  During this work it also became apparent that
the description of nuclear compressibility was unsatisfactory, since the
squeezing of the central density of light nuclei was overpredicted.  The
deficiency was serious because it starts to become important already at
about $A=120$ and becomes even more pronounced for lighter nuclei.  To
account for compressibility effects for light nuclei and for other
higher-order effects an empirical exponential term was
added$\,^{4,5})$.

The additions of the finite-range surface-energy effects and exponential
term to the droplet model$\,^{5})$ resulted in dramatic
improvements in its predictive properties, as summarized in the
discussion of Table A in Ref.$\,^{4})$.  Mass calculations based
on both the FRLDM$\,^{3})$ and the FRDM$\,^{4})$ were
presented in the 1988 review of mass models in Atomic Data and Nuclear
Data Tables.  These calculations also used an improved pairing model
relative to that used in the 1981 work.  In the 1988 results the error
in the FRDM was 8\% lower than that in the FRLDM\@.

However, there were two major unresolved issues in the 1988
calculations.  First, there existed some deficiencies in the pairing
model and the values of the constants that were used.  Second,
$\epsilon_3$ and $\epsilon_6$ shape degrees of freedom were still not
included, so deviations between calculated and measured masses due to
the omission of these shape degrees of freedom were still present.
Extensive investigations of pairing models and their constants have
now been completed and resulted in an improved formulation of the
pairing model$\,^{9})$.  We have now also minimized the
potential energy with respect to $\epsilon_3$ and $\epsilon_6$ shape
degrees of freedom.  An overview of the results has been given in a
paper on Coulomb redistribution effects$\,^{10})$.  The FRDM\@,
which includes Coulomb redistribution effects, is now our preferred
nuclear mass model.

Although the FRDM is now our preferred model,
we also present results  for the FRLDM for comparative purposes
and for use in studies that assume constant nuclear density.
We therefore specify below both models. Because several of the model
constants are determined by least-squares-minimization of the model
error, we start by defining model error.

\subsection{Model error and adjustment \label{2p1} procedure}

In many studies the model error
has been defined as simply the root-mean-square (rms) deviation,
which as usual is given by
\beq
{\rm rms}=
{\left[ \frac{1}{n}\sum_{i=1}^{n}(M^i_{\rm exp} - M^i_{\rm th})^2\right]}
^\frac{1}{2}
\eeq{A}
Here $M^i_{\rm th}$ is the calculated mass for a particular value of the
proton number $Z$ and neutron number $N$, and $M^i_{\rm exp}$
is the corresponding
measured quantity. There are $n$ such measurements for different
$N$ and $Z$\/.
The choice (\ref{A})
is a reasonable definition when all the errors $\sigma^i_{\rm exp}$
associated with the measurements are small compared to the model
error. However, for large $\sigma^i_{\rm exp}$
the above definition is unsatisfactory, since both the theoretical and
experimental errors contribute to the rms deviation. The definition
(\ref{A}) will therefore
always overestimate the intrinsic model error.

When the experimental errors are large,
it is necessary to use an
approach that ``decouples'' the theoretical and
experimental errors from one
another.
This can be accomplished by observing that the calculated masses are
distributed around the {\it true} masses with a standard deviation
$\sigma_{\rm th}$. There exist powerful statistical methods for
determining the intrinsic model error $\sigma_{\rm th}$. The model
error obtained in this way contains no contributions from the
experimental uncertainties $\sigma_{\rm exp}^i$.
To introduce such an error concept a new set of equations for determining model
parameters and error were derived$\,^{3})$
by use of statistical arguments and the maximum-likelihood (ML) method.
Here we generalize from the original assumption$\,^{3})$
$e^i_{\rm th}\in\;$N(0,$\sigma_{\rm th})$ that the  theoretical
error term $e^i_{\rm th}$ is normally distributed with zero mean
deviation from the true mass to
$e^i_{\rm th}\in\;$N($\mu_{\rm th},\sigma_{\rm th})$  to
allow for an error with a mean $\mu_{\rm th}$ that is different from zero
and a standard deviation $\sigma_{\rm th}$ around this
mean$\,^{12})$.
This leads to the generalized  equations
\beq
\sum_{i=1}^{n}\frac{{[M^i_{\rm exp}-(M^i_{\rm th}+{\mu_{\rm th}}^*)]}}
{{\sigma^i_{\rm exp}}^2+{{\sigma _{\rm th}}^2}^*}
\frac{\partial M^i_{\rm th}}{\partial p_{\nu}} =0, \; \; \; \;
\nu=1,2,\ldots, m
\eeq{name12}
\mbox{ }
\beq
\sum_{i=1}^{n}\frac
{{[M^i_{\rm exp}-(M^i_{\rm th}+{\mu_{\rm th}}^*)]}^2
-({\sigma^i_{\rm exp}}^2+{{\sigma_{\rm th}}^2}^*)}
{{({\sigma^i_{\rm exp}}^2+{{\sigma_{\rm th}}^2}^*)}^2}=0
\eeq{deveq13}
\beq
\sum_{i=1}^{n}\frac
{{[M^i_{\rm exp}-(M^i_{\rm th}+{\mu_{\rm th}}^*)]}}
{{({\sigma^i_{\rm exp}}^2+{{\sigma_{\rm th}}^2}^*)}}=0
\eeq{mean1}
where $p_{\rm \nu}$ are the unknown parameters of the model.
The  notation ${{\sigma _{\rm th}}^2}^*$ means that by solving
Eqs.~(\ref{deveq13}) and~(\ref{mean1})
we obtain the estimate ${{\sigma _{\rm th}}^2}^*$ of the
true ${\sigma _{\rm th}}^2$.
Equation (\ref{name12}) is  equivalent to minimizing $S$ with respect
to $p_{\nu}$, where \beq
S=\sum_{i=1}^{n}\frac{{[M^i_{\rm exp}-(M^i_{\rm th}+{\mu_{\rm th}}^*)]}^2}
{{\sigma^i_{\rm exp}}^2+{{\sigma _{\rm th}}^2}^*}
\eeq{deveq14}

Thus, we are led to two additional equations relative to the usual
least-squares equations
that arise when model parameters are estimated by adjustments to
experimental data under the assumption  of a perfect theory,
$\sigma_{\rm th}= 0$. For the
FRLDM the
least-squares equations~(\ref{name12}) are linear, whereas
 for the FRDM they are non-linear.

When the model contains a term $c_0A^0$  that is strictly
constant, Eq.~(\ref{mean1})
is identical to the member in Eq.~(\ref{name12}) that corresponds
to the derivative with respect to this constant.
Thus, one should in this case put ${\mu_{\rm th}}^*=0$ and solve only
the remaining $m+1$ equations.
One may therefore in this case characterize the
error of the model in the region where the parameters were adjusted
solely by the quantity $\sigma_{\rm th}$.
In other cases one should solve the full set of equations. If ${\mu_{\rm
th}}^*$
is significantly different from zero the theory needs modification.
Even if $\mu_{\rm th}=0$ in the original data region,
it is entirely possible (although undesirable) that one obtains a
mean error ${\mu_{\rm th}}^*$ that is substantially different from zero
when one analyzes model
results for new data points to which the parameters were not adjusted.
In this case the most complete characterization of the theoretical error
requires both its
mean $\mu_{\rm th}$ and its standard deviation $\sigma_{\rm th}$ around
this mean.

To allow for a single error measure that is similar to
an rms deviation between the data and model we later also calculate
the square root of the
second central moment of the error term, $\sigma_{{\rm th};\mu=0}$,
in our studies of model behavior in new regions of nuclei.
This quantity is obtained by setting ${\mu_{\rm th}}^*=0$ when solving
Eq.~(\ref{deveq13}). In contrast to the rms measure, it has the
advantage that it has no contributions from the experimental errors.

Equations (\ref{name12})--(\ref{mean1})
constitute a system of $m+2$ equations
that are to be solved together.
It is instructive to rewrite Eqs.~(\ref{deveq13}) and (\ref{mean1}) as
\beq
{{\sigma_{\rm th}}^2}^*=\frac{1}{\sum _{i=1}^n {w_i}^{k_{\sigma}}}
\sum _{i=1}^n {w_i}^{k_{\sigma}}
\left[ (M^i_{\rm exp}-M^i_{\rm th} - {\mu_{\rm th}}^*)^2
-{\sigma ^i_{\rm exp}}^2\right]
\eeq{name16}
\beq
{\mu_{\rm th}}^*=\frac{1}{\sum _{i=1}^n {w_i}^{k_{\mu}}}
\sum _{i=1}^n {w_i}^{k_{\mu}}
\left[ (M^i_{\rm exp}-M^i_{\rm th})\right]
\eeq{mean3}
where
\beq
{w_i}^k=\frac{1}{({\sigma^i_{\rm exp}}^2+{{\sigma _{\rm th}}^2}^*)^k}\\
\eeq{name17}
\beq
k_{\sigma}=2
\eeq{name18}
\beq
k_{\mu} = 1
\eeq{mean4}
The unknowns ${\mu_{\rm th}}^*$ and
${{\sigma _{\rm th}}^2}^*$  can easily be determined
from  Eqs.~(\ref{name16}) and~(\ref{mean3}) by an iterative procedure
whose convergence is extremely
rapid, requiring only about four iterations.
 An {\it interpretation\/}, not a proof, of Eq.~(\ref{name16}) is
that the experimental error is ``subtracted out'' from the
difference between the experimental and calculated masses.

A common misconception is that one has to ``throw away'' data
points
that have errors that are
equal to or larger than the error of the
model whose parameters are determined.
When a proper statistical approach, such as
the one above, is used, this is no longer necessary.

We will see below that the discrepancy between our mass calculations and
measured masses systematically increases as the size
of the nuclear system decreases. It is therefore of interest to consider
that the mass-model error is a function of mass number $A$. A simple
function to investigate is
\beq
\sigma_{\rm th}= \frac{c}{A^{\alpha}}
\eeq{errfunc}
where $c$ and $\alpha$ are two parameters to be determined.
Whereas under the assumption of a constant model error one
determines this single error constant from Eq.~(\ref{name16}),
we find that the ML method for the error assumption in
Eq.~(\ref{errfunc}), with two unknowns,  and assuming $\mu_{\rm th}=0$,
yields the equations
\beq
\sum_{i=1}^{n}\frac
{\begin{displaystyle} {(M^i_{\rm exp}- M^i_{\rm th})}^2
-\left[ {\sigma^i_{\rm exp}}^2
+ {\left( \frac{c^*}{{A_i}^{\alpha^*}} \right) }^2\right]\end{displaystyle}}
{\begin{displaystyle}{\left[{\sigma^i_{\rm exp}}^2
+{\left( \frac{c^*}{{A_i}^{\alpha^*}} \right) }^2\right] }^2
{A_i}^{\alpha^*}\end{displaystyle}}=0
\eeq{deveq21}
\beq
\sum_{i=1}^{n}\frac
{\begin{displaystyle} {(M^i_{\rm exp}- M^i_{\rm th})}^2
-\left[ {\sigma^i_{\rm exp}}^2
+ {\left( \frac{c^*}{{A_i}^{\alpha^*}} \right) }^2\right]\end{displaystyle}}
{\begin{displaystyle}{\left[{\sigma^i_{\rm exp}}^2
+{\left( \frac{c^*}{{A_i}^{\alpha^*}} \right) }^2\right] }^2
{A_i}^{\alpha^*+1}\end{displaystyle}}=0
\eeq{deveq22}
These equations are considerably more complicated to solve than
Eq.~(\ref{name16}).
We solve them by minimizing the sum of the squares of the right members
of Eqs.~(\ref{deveq21}) and (\ref{deveq22}).

\subsection{Shape parameterizations}

The original parameterization of the folded-Yukawa single-particle
model was the three-quadratic-surface
parameterization$\,^{29,32})$. It was
designed to allow great flexibility in describing shapes late
in the fission process. However, it is less suitable for describing
ground-state shapes.

To allow a better description of ground-state shapes and to allow
close comparison with results of Nilsson modified-oscillator
calculations, we incorporated the Nilsson perturbed-spheroid parameterization,
or $\epsilon$ parameterization, into the folded-Yukawa single-particle
computer code in 1973$\,^{30,33,34})$.

In our work here we use the $\epsilon$ parameterization for
all calculations related to ground-state properties.
In our adjustment of macroscopic constants we also
include 28 outer saddle-point heights of fission barriers.
The shapes of these saddle points were obtained in a
three-parameter calculation in the three-quadratic-surface
parameterization in 1973$\,^{33})$.

\subsubsection{Perturbed-spheroid parameterization}

The $\epsilon$
parameterization was originally
used by Nilsson$\,^{35})$ in the
modified-oscillator single-particle potential. It was
introduced to
limit the dimensions of the matrices from which
the single-particle energies and wave functions are
obtained by diagonalization.
This requirement leads to somewhat  complex
expressions for the nuclear shape.
Here we employ its extension to higher-multipole distortions.
For completeness we define it with
axially asymmetric shapes$\,^{36-38})$
included, although this symmetry-breaking shape
degree of freedom has
not yet been implemented in the folded-Yukawa single-particle
model.
Note that the factor $\sqrt{\frac{4\pi}{9}}\frac{1}{2}$ is
missing in front of the $V_4(\gamma)$ function in
Eq.~(3) of Ref.$\,^{38})$.

As the  first step
in defining the $\epsilon$ parameterization
a ``stretched'' representation is introduced.
The stretched coordinates $\xi$, $\eta$, and $\zeta$ are defined
by
\beqar
\xi & = & {\left\{ \frac{m\omega_0}{\hbar}\left[1
       -\frac{2}{3}\epsilon_2\cos
        \left(\gamma+\frac{2}{3}\pi\right)\right]\right\}
}^{1/2}x
        \nonumber  \\[1ex]
\eta & = & {\left\{ \frac{m\omega_0}{\hbar}\left[1
       -\frac{2}{3}\epsilon_2\cos
        \left(\gamma-\frac{2}{3}\pi\right)\right]\right\}
}^{1/2}y
      \nonumber     \\[1ex]
\zeta & = & {\left\{ \frac{m\omega_0}{\hbar}\left[1
       -\frac{2}{3}\epsilon_2\cos
        \gamma\right]\right\} }^{1/2}z
\eeqar{stretched}
where $\hbar\omega_0$ is the
oscillator
energy, $\epsilon_2$ the ellipsoidal deformation parameter, and
$\gamma$ the
non-axiality angle. It is then convenient to define a
``stretched'' radius vector
$\rho_{\rm t}$   by
\beq
\rho_{\rm t} = (\xi^2 + \eta^2 + \zeta^2)^{1/2}
\eeq{rhot}
a stretched polar angle  $\theta_t$ by
\beq
u = \cos \theta_t = \frac{\zeta}{\rho_{\rm t}}=
\left[ \frac{
\begin{displaystyle}
1-\frac{2}{3}\epsilon_2\cos \gamma
\end{displaystyle}  }
{\begin{displaystyle}
1-\frac{1}{3}\epsilon_2\cos\gamma(3\cos^2\theta-1)
+\left(\frac{1}{3}\right)^{1/2}\epsilon_2\sin\gamma\sin^2\theta\-
\cos2\phi
\end{displaystyle}}
\right]^{1/2} \cos \theta
\eeq{polart}
and a stretched azimuthal angle $\phi_{\rm t}$ by
\beq
v = \cos2\phi_{\rm t}=\frac{2\eta}{(\xi^2+\eta^2)^{1/2}} =
\frac{\begin{displaystyle}
\left[1+\frac{1}{3}\epsilon_2\cos\gamma\right]\cos2\phi
+\left(\frac{1}{3}\right)^{1/2}\epsilon_2\sin \gamma
\end{displaystyle}}
{\begin{displaystyle}
1+\frac{1}{3}\epsilon_2\cos \gamma
  +\left(\frac{1}{3}\right)^{1/2}\epsilon_2\sin\gamma\cos 2\phi
\end{displaystyle}}
\eeq{azimutht}

In the folded-Yukawa model the single-particle potential
is very different from that in the Nilsson modified-oscillator model.
However, the definition of  the $\epsilon$ parameterization
will be most clear if we follow the steps in the Nilsson model.
The implementation in the folded-Yukawa model will then be simple.
The Nilsson modified-oscillator potential is defined by
\beqar
\lefteqn{
 V = \frac{1}{2}\hbar\omega_0 {\rho_{\rm t}}^2\left\{ 1
         + 2 \epsilon_1 P_1(\cos\theta_{\rm t})
                 \phantom{\frac{1}{3}} \right. }
                                    \nonumber \\[1ex]
& &  \phantom{\frac{1}{3}}
      - \frac{2}{3} \epsilon_2 \cos\gamma P_2(\cos\theta_{\rm t})
      + \frac{1}{3}\epsilon_2 \sin \gamma
\left(\frac{8}{5}\pi\right)^{1/2}
        \left[Y_2^2(\theta_{\rm t},\phi_{\rm t})
             +Y_{2}^{-2}(\theta_{\rm t},\phi_{\rm t})\right]
                \nonumber \\[1ex]
& & \left. \phantom{\frac{1}{3}}
    + 2 \epsilon_3 P_3(\cos \theta_{\rm t})
    + 2 \epsilon_4 V_4(\cos \theta_{\rm t},\cos 2\phi_{\rm t})
    + 2 \epsilon_5 P_5(\cos \theta_{\rm t})
    + 2 \epsilon_6 P_6(\cos \theta_{\rm t})\right\} \nonumber
\\[1ex]
& &  \phantom{\frac{1}{3}}
       -\kappa \hbar \! \stackrel{\circ}{\omega}_0 \! \!
       \left[ 2\vec{l}_{\rm t}\cdot\vec{s} +
       \mu({\vec{l}_{\rm t}}^{\phantom{2}2} -
      <{\vec{l}_{\rm t}}^{\phantom{2}2}>)\right]
\eeqar{vosc}
where
\beq
V_4(u,v) =a_{40}P_4 +
\sqrt{\frac{\begin{displaystyle} 4\pi \end{displaystyle}}
{\begin{displaystyle} 9 \end{displaystyle}}}
\left[ a_{42}(Y_4^2 + Y_4^{-2}) + a_{44}(Y_4^4 + Y_4^{-4})\right]
\eeq{v4}
Here the hexadecapole potential $V_4(u,v)$ is made dependent
on $\gamma$ in such a way that axial symmetry is maintained
when $\gamma=0$, $60^{\circ}$, $-120^{\circ}$, and $-60^{\circ}$,
for mass-symmetric shapes and for $\epsilon_6=0$.
This is accomplished by choosing the coefficients $a_{4i}$ so
that they have the transformation properties of a hexadecapole
tensor.  However, this is achieved only for mass-symmetric
shapes and for $\epsilon_6=0$. The $\epsilon$
parameterization has not been generalized to a more
general case. Thus$\,^{38})$
\beqar
a_{40}&=&  \frac{1}{6}(5\cos^2\gamma + 1) \nonumber \\[1ex]
a_{42}&=&  -\frac{1}{12}\sqrt{30}\sin 2 \gamma  \nonumber \\[1ex]
a_{44}&=& \frac{1}{12}\sqrt{70}\sin^2\gamma
\eeqar{a4cof}

It is customary to now assume that the shape of the
nuclear
surface is equal to the shape of an equipotential surface given
by Eq.~(\ref{vosc}). By neglecting the $\vec{l}_{\rm
t}\cdot\vec{s}$
and ${\vec{l}_{\rm t}}^{\phantom{2}2}$ terms and solving for
$\rho_{\rm t}$
and then using Eqs.~(\ref{stretched})--(\ref{azimutht}) to derive
an expression for $r$ in the non-stretched laboratory system
we obtain
\beqar
r(\theta,\phi)& = &
\frac{R_0}{\omega_0/ \! \stackrel{\circ}{\omega}_0}
 \left\{
 \left[1-\frac{2}{3}\epsilon_2
\cos\left(\gamma+\frac{2}{3}\pi\right)\right]
 \left[1-\frac{2}{3}\epsilon_2
\cos\left(\gamma-\frac{2}{3}\pi\right)\right]
 \left[1-\frac{2}{3}\epsilon_2 \cos\gamma\right] \right\}^{-1/2}
                            \nonumber  \\[1ex]
  & & \phantom{xxx} \times \left[
   1-\frac{1}{3}\epsilon_2 \cos \gamma -\frac{2}{9}{\epsilon_2}^2
   \cos^2\gamma +\epsilon_2\left(\cos \gamma
  + \frac{1}3\epsilon_2\cos 2\gamma\right)u^2
       \phantom{\left(\frac{1}{3}\right)^{1/2}}\right.
      \nonumber \\[1ex]
 & & \phantom{xxxxxxxxxxxxx}  \left.
        -\left(\frac{1}{3}\right)^{1/2}\epsilon_2\sin \gamma
 \left(1-\frac{2}{3}\epsilon_2\cos
\gamma\right)(1-u^2)v\right]^{1/2}
      \nonumber \\[1ex]
 & & \phantom{xxx} \times \left[ 1 - \frac{2}{3}\epsilon_2 \cos
\gamma\frac{1}{2}(3u^2-1)
+\left(\frac{1}{3}\right)^{1/2}\epsilon_2 \sin \gamma(1-u^2)v
\right.
\nonumber \\[1ex]
 & & \phantom{x} \left. \phantom{\left(\frac{1}{3}\right)^{1/2}}
    + 2 \epsilon_1 P_1(u)
    + 2 \epsilon_3 P_3(u)
    + 2 \epsilon_4 V_4(u,v)
    + 2 \epsilon_5 P_5(u)
    + 2 \epsilon_6 P_6(u)\right]^{-1/2}
\eeqar{radiusv}

In the Nilsson model the starting point is to define the potential. After
the potential has been generated the shape of the nuclear
surface is deduced by the above argument. In the folded-Yukawa
model the starting point is different. There, the
equation for
the nuclear surface, given by Eq.~(\ref{radiusv}) in the case of the
$\epsilon$ parameterization, is
specified in the initial step. Once the
shape of the surface is known, the single-particle potential
may be generated as described in later sections.

The quantity $\omega_0/ \! \stackrel{\circ}{\omega}_0$ is determined
by requiring that
the volume remain constant with deformation, which gives
\beqar
\left( \frac{\omega_0}{\stackrel{\circ}{\omega}_0}\right)^3
& = & \frac{1}{4\pi}
 \left\{
 \left[1-\frac{2}{3}\epsilon_2
\cos\left(\gamma+\frac{2}{3}\pi\right)\right]
 \left[1-\frac{2}{3}\epsilon_2
\cos\left(\gamma-\frac{2}{3}\pi\right)\right]
 \left[1-\frac{2}{3}\epsilon_2 \cos\gamma\right] \right\}^{-1/2}
                            \nonumber  \\[1ex]
  & & \times \int_0^{\pi}d\theta_{\rm
t}\int_0^{2\pi}d\phi_{\rm t}\sin\theta_{\rm t}
\left[ 1 - \frac{2}{3}\epsilon_2 \cos \gamma P_2(u)
+\epsilon_2\sin\gamma\left(\frac{8\pi}{45}\right)^{1/2}
(Y_{2}^2+Y_{2}^{-2})\right.
\nonumber \\[1ex]
 & & \phantom{x} \left. \phantom{\left(\frac{1}{3}\right)^{1/2}}
    + 2 \epsilon_1 P_1(u)
    + 2 \epsilon_3 P_3(u)
    + 2 \epsilon_4 V_4(u,v)
    + 2 \epsilon_5 P_5(u)
    + 2 \epsilon_6 P_6(u)\right]^{-3/2}
\eeqar{volcon}
The above equation is derived by determining the volume inside
the nuclear surface given by Eq.~(\ref{radiusv}), with  the
integral
$\int d^3r$ inside the surface evaluated
in terms of
the ``non-stretched'' coordinates $\theta$ and $\phi$.
After
a variable substitution one arrives at the expression in
Eq.~(\ref{volcon}).

The Legendre polynomials
$P_l$ occurring in the
definitions of the $\epsilon$ parameterization are defined by
\beq
P_l(u)=\frac{1}{2^ll!}\frac{d^l}{du^l}(u^2-1)^l \; \;,
\; \; \; \; \; l=0,\; 1,\; 2,\ldots,\infty
\eeq{legdef}
The first six Legendre polynomials are
\beqar
P_0(u) & = & 1                                           \nonumber \\[1ex]
P_1(u) & = & u                                           \nonumber \\[1ex]
P_2(u) & = & \frac{1}{2}(3u^2-1)                         \nonumber \\[1ex]
P_3(u) & = & \frac{1}{2}(5u^3-3u)                        \nonumber \\[1ex]
P_4(u) & = & \frac{1}{8}(35u^4 - 30 u^2 +3)              \nonumber \\[1ex]
P_5(u) & = & \frac{1}{8}(63u^5 - 70u^3 + 15u)            \nonumber \\[1ex]
P_6(u) & = & \frac{1}{16}(231u^6 - 315u^4 + 105u^2 - 5)
\eeqar{legpol}

The associated Legendre functions
$P_l^m$
are defined by
\beq
P_l^m(u)=\frac{(1-u^2)^{m/2}}{2^ll!}\frac{d^{l+m}}{du^{l+m}}(u^2-1)^l \; \;
, \; \; \; \; \; l=0,\; 1,\; 2,\ldots,\infty; \; \; \;
                      m=0,\; 1,\; 2,\ldots,l
\eeq{aslegdef}
The spherical harmonics are  then determined from the  relations
\beq
Y_l^m(\theta,\phi) = (-)^m\left[\frac{(2l+1)}{4\pi}
\frac{(l-m)!}{(l+m)!}\right]^{1/2}P_l^m(\cos\theta)e^{im\phi}\; \;,
\; \; \; \; m\geq 0
\eeq{ylmdef}
\beq
{Y_l^m}^*(\theta,\phi) = (-)^mY_l^{-m}(\theta,\phi)
\eeq{conjug}
which yield for the functions used here
\beqar
Y_2^2(\theta,\phi) & = & \sqrt{\frac{15}{32\pi}}\sin^2\theta e^{2i\phi}
  \nonumber \\[1ex]
Y_2^{-2}(\theta,\phi) & = & \sqrt{\frac{15}{32\pi}}\sin^2\theta e^{-2i\phi}
   \nonumber \\[1ex]
Y_4^{4}(\theta,\phi) & = & \sqrt{\frac{315}{512\pi}}\sin^4\theta e^{4i\phi}
   \nonumber \\[1ex]
Y_4^{-4}(\theta,\phi) & = & \sqrt{\frac{315}{512\pi}}\sin^4\theta e^{-4i\phi}
   \nonumber \\[1ex]
Y_4^{2}(\theta,\phi) & = & \sqrt{\frac{45}{128\pi}}\sin^2\theta
(7\cos^2\theta - 1)e^{2i\phi}
   \nonumber \\[1ex]
Y_4^{-2}(\theta,\phi) & = & \sqrt{\frac{45}{128\pi}}\sin^2\theta
(7\cos^2\theta - 1)e^{-2i\phi}
\eeqar{ylm}
The sums
\beqar
SY_{22} & = & Y_2^2(\theta,\phi) + Y_2^{-2}(\theta,\phi)
\nonumber \\[1ex]
SY_{44} & = & Y_4^4(\theta,\phi) + Y_4^{-4}(\theta,\phi)
 \nonumber \\[1ex]
SY_{42} & = & Y_4^2(\theta,\phi) + Y_4^{-2}(\theta,\phi)
\eeqar{fydef}
 are required in the expression for the single-particle potential and
in the corresponding equation for the nuclear surface.
We obtain
\beqar
SY_{22}
& = & \sqrt{\frac{15}{8\pi}}\sin^2\theta \cos 2\phi =
\sqrt{\frac{15}{8\pi}}(1-u^2)v
\nonumber \\[1ex]
SY_{44}
& = & \sqrt{\frac{315}{128\pi}}\sin^4\theta \cos 4\phi =
\sqrt{\frac{15}{128\pi}}(1-u^2)^2(2v^2 -1)
 \nonumber \\[1ex]
SY_{42}
& = &
\sqrt{\frac{45}{32\pi}}\sin^2\theta (7\cos^2\theta-1)\cos 2\phi =
\sqrt{\frac{45}{32\pi}}(1-u^2)(7u^2 -1)v
\eeqar{sumylm}

\subsubsection{Three-quadratic-surface parameterization}

In the
three-quadratic-surface parameterization  the shape of the nuclear surface
is specified in terms of three smoothly joined portions of quadratic surfaces
of revolution. They are completely specified by $\,^{32})$
\beqar
\rho^2= \left\{ \begin{array}{ll}
{a_1}^2-
\begin{displaystyle}\frac{{a_1}^2}{{c_1}^2}\end{displaystyle}
(z-l_1)^2 \; \; ,&l_1-c_1\leq z \leq z_1\\[2ex]
{a_2}^2-
\begin{displaystyle}\frac{{a_2}^2}{{c_2}^2}\end{displaystyle}
(z-l_2)^2 \; \; ,&z_2 \leq z \leq l_2+c_2 \\[2ex]
{a_3}^2-
\begin{displaystyle}\frac{{a_3}^2}{{c_3}^2}
\end{displaystyle}
(z-l_3)^2 \; \; ,& z_1 \leq z \leq z_2
\end{array}
\right.
\eeqar{threeqs}
Here the left-hand surface is denoted by the subscript 1, the right-hand one
by 2, and the middle one by 3.
Each surface is specified by the position $l_i$ of
its center, its transverse semiaxis $a_i$, and
its semi-symmetry axis $c_i$.
At the
left and right intersections of the middle surface with the end surfaces
the value of $z$ is $z_1$ and $z_2$, respectively.

There are nine numbers required to specify the expressions in Eq.\
(\ref{threeqs}) but three numbers are eliminated by
the conditions of constancy of the volume and continuous
first derivatives at $z_1$ and $z_2$.
The introduction of  an auxiliary unit of distance $u$ through
\beq
u=\left[\frac{1}{2}\left({a_1}^2+{a_2}^2\right) \right] ^{\frac{1}{2}}
\eeq{unitl}
permits a natural definition of two sets of shape coordinates. We define
three mass-symmetric coordinates $\sigma _i$ and three
mass-asymmetric coordinates
$\alpha _i$ by
\beqar
 \sigma_1&=&\begin{displaystyle}\frac{(l_2-l_1)}{u}
\end{displaystyle}  \nonumber \\[2ex]
 \sigma_2&=&\begin{displaystyle}\frac{{a_3}^2}{{c_3}^2}
\end{displaystyle}  \nonumber \\[2ex]
 \sigma_3&=&\begin{displaystyle}\frac{1}{2}\left(\frac{{a_1}^2}{{c_1}^2}+
\frac{{a_2}^2}{{c_2}^2}\right)
\end{displaystyle}  \nonumber \\[2ex]
 \alpha_1&=&\begin{displaystyle}\frac{1}{2}\frac{(l_1+l_2)}{u}
\end{displaystyle}  \nonumber \\[2ex]
 \alpha_2&=&\begin{displaystyle}\frac{({a_1}^2-{a_2}^2)}{u^2}
\end{displaystyle}  \nonumber \\[2ex]
\alpha_3&=&\begin{displaystyle}\frac{{a_1}^2}{{c_1}^2}-\frac{{a_2}^2}{{c_2}^2}
\end{displaystyle}
\eeqar{coor}
The coordinate $\alpha_1$ is not varied freely but is instead determined by
the requirement that the center of mass be at the origin.

\subsubsection{Conversions to $\beta$ parameters}

A common parameterization, which we do {\it not} use here,
is the $\beta$ parameterization. However, since we
want to present some of our results in terms of
$\beta$ shape parameters, we introduce the parameterization
and a scheme to express shapes generated in other
parameterizations in terms of $\beta$ deformation parameters.
In the $\beta$ parameterization the radius vector $r$ is
defined by
\beq
r(\theta,\phi) = R_0(1 + \sum_{l=1}^{\infty} \sum_{m= -l}
^{l}\beta_{lm}Y_{l}^m)
\eeq{betapar}
where $R_0$ is deformation dependent so as to
conserve the volume inside the nuclear surface.
When only axially symmetric shapes are considered the
notation $\beta_l$ is normally used for $\beta_{l0}$.
Since the spherical harmonics $Y_{l}^m$ are
orthogonal, one may determine the
$\beta$ parameters corresponding to a specific shape
in the $\epsilon$ parameterization  by use of
\beq
\beta_{lm}= \sqrt{4 \pi} \frac{
\begin{displaystyle}
\int r(\theta,\phi)Y_{l}^m(\theta,\phi)d\Omega
\end{displaystyle} }
{ \begin{displaystyle}
\int r(\theta,\phi)Y_{0}^0(\theta,\phi)d\Omega
\end{displaystyle} }
\eeq{betaconv}
where $r$ is now the radius vector in the $\epsilon$ parameterization,
given by Eq.~(\ref{radiusv}). This conversion equation is in fact valid
for a radius vector $r(\theta,\phi)$ defined by any parameterization.

When the $\beta$ parameters corresponding to a specific shape in the
$\epsilon$ parameterization are determined one should observe that
higher-order $\beta$ parameters may be non-zero even if higher-order
$\epsilon$ parameters are identically zero. For this reason, and
because $\beta_5$ is not tabulated, the nuclear ground-state shape is
not completely specified by the $\beta$ parameters in the Table,
whereas the shape is completely defined by the $\epsilon$ parameters.

\subsection{Finite-range droplet model}

The {\it finite-range droplet model},
developed in 1984$\,^{5})$, combines
the finite-range effects of the FRLDM$\,^{31,39,40})$
with the higher-order terms in
the droplet model.
In addition, the finite-range
droplet model contains the new exponential term
\beq
-CAe^{-\gamma A^{1/3}}\overline{\epsilon}
\eeq{expterm}
where $C$ and $\gamma$ specify the strength and range, respectively,
of this contribution to the energy and
the quantity $\overline{\epsilon}$ is a dilatation variable given
by Eq.~(\ref{name6}) below.
The exponential term leads to an improved
description of compressibility effects
and is crucial to the substantially improved results obtained in the
finite-range droplet model relative to the original droplet model.
The necessity for this empirical exponential term,
which is discussed extensively in
Refs.$\,^{5,41})$, is clearly demonstrated in
Refs.$\,^{5,41})$
and by the results obtained in Sec.\ 3.2 below.

Most of our results here are based on the finite-range droplet model
for the macroscopic term. Relative to the formulation
given in Ref.$\,^{5})$, which unfortunately has numerous
misprints, we use a new model for the average
neutron and proton pairing gaps.
The complete expression for the
contribution to the atomic mass excess from the FRDM macroscopic energy
is obtained after minimization with
respect to variations in $\overline{\epsilon}$ and $\overline{\delta}$,
where $\overline{\delta}$ is the average bulk relative neutron excess
given by Eq.~(\ref{name8}) below. One then obtains
\newpage \vspace{-0.4in}
\begin{eqnarray}
\lefteqn{E_{\rm mac}(Z,N,{\rm shape}) =} \nonumber \\[2ex]
& &\begin{array}{rclr}
& &\begin{displaystyle}  M_{\rm H}Z+M_{\rm n}N \end{displaystyle}
 & \mbox{  \phantom{mass}mass excesses of $Z$ hydrogen atoms
and $N$ neutrons}
\nonumber \\[3ex]
&+& \lefteqn{ \begin{displaystyle} \left(-a_1 + J
\overline{\delta} ^2 - \frac{1}{2}K
\overline{\epsilon} ^2\right)A \end{displaystyle} }
 & \mbox{  volume energy} \nonumber \\[4ex]
&+&\lefteqn{ \begin{displaystyle}
\left(a_2B_1 + \frac{9}{4}\frac{J^2}{Q}\overline{\delta} ^2
\frac{{B_{\rm s}}^2}{B_1}\right)
A^{2/3} \end{displaystyle}}
 & \mbox{  surface energy}\nonumber \\[4ex]
&+&a_3A^{1/3}B_{\rm k}
 & \mbox{  curvature energy}\nonumber \\[3ex]
&+&a_0A^0
 & \mbox{   $A^0$ energy} \nonumber \\[3ex]
&+&\begin{displaystyle} c_1\frac{Z^2}{A^{1/3}}B_3
\end{displaystyle}
 & \mbox{  Coulomb energy} \nonumber \\[4ex]
&-&c_2Z^2A^{1/3}B_{\rm r}
 & \mbox{  volume redistribution energy} \nonumber \\[3ex]
&-&\begin{displaystyle} c_4\frac{Z^{4/3}}{A^{1/3}}
\end{displaystyle}
 & \mbox{  Coulomb exchange correction} \nonumber \\[4ex]
&-&\begin{displaystyle} c_5Z^2\frac{B_{\rm w}B_{\rm s}}{B_1}
\end{displaystyle}
 & \mbox{\hfill surface redistribution energy} \nonumber \\[4ex]
&+&\begin{displaystyle}f_0\frac{Z^2}{A}
\end{displaystyle}
 & \mbox{\hfill proton form-factor correction to the Coulomb energy} \nonumber
\\[4ex]
&-&c_{\rm a}(N-Z)
 & \mbox{\hfill charge-asymmetry energy} \nonumber \\[4ex]
&+&\lefteqn{ W \left( |I|+ \left\{ \begin{array}{ll} 1/A \; \; ,
 & \mbox{$Z$ and $N$ odd and
 equal} \\
 0 \; \; ,& {\rm otherwise}
\end{array}  \right. \right) }
&\mbox{Wigner energy}
\nonumber \\[5ex]
&+ & \lefteqn{ \left\{ \begin{array}{ll}
+\begin{displaystyle}
\; \overline{\Delta} _{\rm p} + \overline{\Delta} _{\rm n} -
\delta _{\rm np}\; \; ,
\end{displaystyle}
&   \mbox{$Z$ {\rm and} $N$ {\rm odd}} \\[2ex]
+\begin{displaystyle}
\; \overline{\Delta} _{\rm p}  \; \; ,
\end{displaystyle}
&   \mbox{$Z$ {\rm odd and} $N$ {\rm even}} \\[2ex]
+\begin{displaystyle} \; \overline{\Delta} _{\rm n} \; \; , \end{displaystyle}
&
  \mbox{$Z$ {\rm even and}  $N$ {\rm odd}} \\[2ex]
+\begin{displaystyle}
\; 0 \; \; ,
\end{displaystyle} &
  \mbox{$Z$ {\rm and} $N$ {\rm even}}
\end{array} \right. }
& \mbox{ average pairing energy} \\[12ex]
&-&a_{\rm el}Z^{2.39}
 & \mbox{\hfill energy of bound electrons}
\end{array}  \nonumber \\
\eeqar{macener}
This expression differs from the corresponding one used in our earlier
calculations$\,^{5})$ only in the form of the
average pairing energy appearing in the next-to-last term.
One should note that after minimization the exponential term is
present only implicitly in Eq.~(\ref{macener}) through its
presence in Eq.~(\ref{name6}) below.
For the average neutron pairing gap
$\overline{\Delta} _{\rm n}$, average proton pairing gap
$\overline{\Delta} _{\rm p}$,
and
average neutron-proton interaction energy $\delta _{\rm np}$
we now
use$\,^{9,24,42})$
\beq
\overline{\Delta}_{\rm n}=\frac{r_{\rm mac}B_{\rm s}}{N^{1/3}}
\eeq{pairn}
\beq
\overline{\Delta}_{\rm p}=\frac{r_{\rm mac}B_{\rm s}}{Z^{1/3}}
\eeq{pairp}
\beq
\delta _{\rm np} = \frac{h}{B_{\rm s}A^{2/3}}
\eeq{pairnp}
The zero reference point for the pairing energy now corresponds to
even-even nuclei rather than to halfway between even-even and odd-odd
nuclei.

The quantities $c_1$, $c_2$, $c_4$, and $c_5$ are defined by
\beqar
c_1&=&\begin{displaystyle}
\frac{3}{5}\frac{e^2}{r_0}
\end{displaystyle}\nonumber  \\[2ex]
c_2&=&\begin{displaystyle}
\frac{1}{336}\left(\frac{1}{J}+\frac{18}{K}\right){c_1}^2
\end{displaystyle}\nonumber  \\[2ex]
c_4&=&\begin{displaystyle}
\frac{5}{4}{\left(\frac{3}{2\pi}\right)}^{2/3}c_1
\end{displaystyle}\nonumber  \\[2ex]
c_5&=&\begin{displaystyle}
\frac{1}{64Q}{c_1}^2
\end{displaystyle}
\eeqar{name2a}

In Eq.~(\ref{macener}) we have kept only the first term
in the expression for the proton form-factor
correction to the Coulomb energy, so that $f_0$ is
given by
\beq
f_0=  -\frac{1}{8}\left( \frac{145}{48} \right) \frac{{r_{\rm
p}}^2e^2}{{r_0}^3}\\[2ex]
\eeq{name3}
The bulk nuclear asymmetry $\delta$ is defined
in terms of the neutron density $\rho_{\rm n}$ and
proton density $\rho_{\rm p}$ by
\beq
\delta = \frac{\rho_{\rm n}- \rho_{\rm p}}{\rho_{\rm bulk}}
\\[2ex] \eeq{name7}
and the $average$ bulk nuclear asymmetry is given by
\beq
\overline{\delta} =
\left(I+\frac{3}{16}\frac{c_1}{Q}\frac{Z}{A^{2/3}}
\frac{B_{\rm v}B_{\rm s}}{B_1}\right)/
\left(1+\frac{9}{4}\frac{J}{Q}\frac{1}{A^{1/3}}\frac{{B_{\rm s}}^2}{B_1}\right)
\\[2ex] \eeq{name8}

The relative deviation in the bulk of the density $\rho$
from its nuclear matter value $\rho_0$ is defined by
\beq
\epsilon =  -\frac{1}{3}\frac{\rho - \rho_0}{\rho_0} \\[2ex]
\eeq{name5}
and the $average$ relative deviation in the bulk of the density is given
by
\beq
\overline{\epsilon} = \left(C e^{-\gamma A^{1/3}}-2a_2\frac{B_2}{A^{1/3}}
+L\overline{\delta} ^2 + c_1\frac{Z^2}{A^{4/3}}B_4\right)/K \\[2ex]
\eeq{name6}
The quantity $B_1$ is the relative generalized surface or nuclear  energy in a
model that accounts for the effect of the finite range of the nuclear
force. It is given by
\beq
B_1=\frac{A^{-2/3}}{8 \pi ^2 {r_0}^2a^4}
\int \! \!\int_{V} \left( 2-\frac{|{\rm {\bf r}}-{\rm {\bf r}}'|}{a}\right)
\frac{e^{-|{\rm {\bf r}}-{\rm {\bf r}}'|/a}}
{|{\rm {\bf r}}-{\rm {\bf r}}'|/a} {d^3r}\,{d^3r'}
\eeq{nucen}
where the integration is over the specified sharp-surface deformed
{\it generating} shape of
volume $V$. Since the volume of the generating shape  is
conserved during deformation
we have
\beq
V=\frac{4\pi }{3}{R_0}^3
\eeq{volume}
where $R_0$ is the radius of the  spherical shape.
The relative Coulomb energy $B_3$ is given by
\beq
B_3 =\frac{15}{32 \pi ^2} \frac{A^{-5/3}}{{r_0}^5}
\int \! \! \int_{V} \frac{{d^3r}\,{d^3r'}}{|{\rm {\bf r}}-{\rm {\bf r}}'|}
\left[ 1- \left( 1 + \frac{1}{2}
\frac{|{\rm {\bf r}}-{\rm {\bf r}}'|}
{a_{\rm den}} \right)
e^{-|{\rm {\bf r}}-{\rm {\bf r}}'|/a_{\rm den}} \right]
\eeq{coulen}

The quantities $B_1$ and $B_3$ are evaluated for $R_0=r_0A^{1/3}$.
However,
in the finite-range droplet model the equilibrium value $R_{\rm den}$
of the
equivalent-sharp-surface radius
corresponding to the nuclear density
is given by the expression
\beq
R_{\rm den}=r_0A^{1/3}(1+\overline{\epsilon})
\eeq{radius}
Thus, the actual value of the nuclear radius is determined by the
balance between Coulomb, compressibility, and surface-tension effects as
expressed by Eq.~(\ref{name6}).  To calculate this balance it is
necessary to know the response of the surface-energy and Coulomb-energy
terms $B_1$ and $B_3$ to size changes.  To account for this response we
introduce the quantities $B_2$ and $B_4$, which are related to the
derivatives of $B_1$ and $B_3$.  These derivatives are evaluated
numerically and during this evaluation the radius $R$ of the {\it
generating} shape is varied around the value $r_0A^{1/3}$.

The quantity $B_2$, which as mentioned above is related
to the derivative of the relative generalized surface energy
$B_1$, is defined by
\beq
B_2=\frac{1}{2x_0}\left[\frac{d}{dx}\left( x^2B_1\right)\right]_{x=x_0}
\\[2ex] \eeq{name10}
with
\beq
x=\frac{R}{a} \; \; \; {\rm and} \; \; \; x_0=\frac{r_0A^{1/3}}{a}
\\[2ex] \eeq{name11}
The quantity $B_4$ is related to the derivative of the relative
Coulomb energy
$B_3$ and is defined by
\beq
B_4= -{y_0}^2 \left[ \frac{d}{dy} \left( \frac{B_3}{y} \right) \right] _{y=y_0}
\\[2ex] \eeq{name13}
with
\beq
y=\frac{R}{a_{\rm den}} \; \; \; {\rm and} \; \; \;
y_0=\frac{r_0A^{1/3}}{a_{\rm den}}
\\[2ex]  \eeq{name14}

For spherical shapes the quantities $B_1$, $B_2$, $B_3$, and $B_4$
can be evaluated
analytically. One obtains
\beqar
B_1^{(0)}& = & 1-\frac{3}{{x_0}^2}+ \left( 1+x_0\right) \left( 2+\frac{3}{x_0}
+\frac{3}{{x_0}^2} \right) e^{-2x_0}
\nonumber \\[2ex]
B_2^{(0)}& = &1-\left(1+2x_0-{x_0}^2\right)e^{-2x_0}
\nonumber \\[2ex]
B_3^{(0)}& = & 1-\frac{5}{{y_0}^2}
\left[1-\frac{15}{8y_0}+\frac{21}{8{y_0}^3}
-\frac{3}{4}\left(1+\frac{9}{2y_0}+
\frac{7}{{y_0}^2}+\frac{7}{2{y_0}^3}
\right) e^{-2y_0} \right ]
\nonumber \\[2ex]
B_4^{(0)}& = &  1+5 \left[ -\frac{3}{{y_0}^2}+\frac{15}{2{y_0}^3}-
\frac{63}{4{y_0}^5}+\frac{3}{4} \left( \frac{2}{y_0}
+\frac{12}{{y_0}^2}
+\frac{32}{{y_0}^3}+\frac{42}{{y_0}^4}+\frac{21}{{y_0}^5}
 \right)e^{-2y_0} \right]
\eeqar{name19}

The expression $B_3$ for the relative
Coulomb energy yields the energy for an
arbitrarily
shaped, homogeneously charged, diffuse-surface nucleus to all orders in
the diffuseness constant $a_{\rm den}$. The constants in front of
the integrals for $B_1$
and $B_3$ are chosen so that $B_1$ and $B_3$ are 1 for a sphere in
the limit in which the range constant $a$ and the diffuseness constant
$a_{\rm den}$ are zero, in analogy with the definition of the quantities
$B_{\rm s}$ and $B_{\rm C}$ in the standard liquid-drop
and droplet models. The quantities $B_2$ and $B_4$, which are related to the
derivatives of $B_1$ and $B_3$, respectively, were introduced above to
treat the response of the nucleus to a change in size,
resulting from a finite
compressibility. The shape-dependent quantities
$B_{\rm s}$, $B_{\rm v}$, $B_{\rm w}$, $B_{\rm k}$, and $B_{\rm r}$,
which are defined$\,^{7})$ in the standard droplet model,
are given by
\beqar
\begin{array}{lr}
B_{\rm s} =
\begin{displaystyle}
\phantom{-}\; \; \frac{A^{-2/3}}{4\pi {r_0}^2}  \int _S dS
\end{displaystyle}
& {\rm surface \; \;energy} \\[3ex]
B_{\rm v}  =
\begin{displaystyle}
 - \frac{15A^{-4/3}}{16\pi^2 {r_0}^2}
\int \! \! \int_{V} \left( \frac{1}
{|{\rm {\bf r}}-{\rm {\bf r}}'|} - \overline{W} \right)
\, d^3r \, d^3r'
\end{displaystyle}
& {\rm neutron \; \; skin \; \; energy}  \\[3ex]
B_{\rm w}  =
\begin{displaystyle}
\phantom{-}\; \; \frac{225A^{-2}}{64\pi^3 {r_0}^6}
\int \! \! \int_{S} {\left( \frac{1}
{|{\rm {\bf r}}-{\rm {\bf r}}'|} - \overline{W} \right) }^2
\, dS \, dS'
\end{displaystyle}
& {\rm surface \; \; redistribution \; \; energy}  \\[3ex]
B_{\rm k}  =
\begin{displaystyle}
\phantom{-}\; \; \frac{A^{-1/3}}{8\pi {r_0}}
\int_{S} \left( \frac{1}{R_1} + \frac{1}{R_2} \right)
\,  dS
\end{displaystyle}
& {\rm curvature \; \; energy}  \\[3ex]
B_{\rm r}  =
\begin{displaystyle}
\phantom{-}\; \; \frac{1575A^{-7/3}}{64\pi^3 {r_0}^7}
\int \! \! \int_{V} {\left( \frac{1}
{|{\rm {\bf r}}-{\rm {\bf r}}'|} - \overline{W} \right) }^2
\, d^3r \, d^3r'
\end{displaystyle}
& {\rm volume \; \; redistribution \; \; energy}
\end{array} \nonumber  \\
\eeqar{dropshap}
where
\beq
\overline{W} =
\begin{displaystyle}
\frac{3 A^{-1}}{4\pi {r_0}^3}
\int \! \! \int_{V}  \frac{1}{|{\rm {\bf r}}-{\rm {\bf r}}'|}
\, d^3r \, d^3r'
\end{displaystyle}
\eeq{avcoul}
is the average of the Coulomb potential and $R_1$ and $R_2$ are
the principal radii of curvature.

\subsection{Values of FRDM macroscopic-model constants}

The constants appearing in the expression for the finite-range droplet
macroscopic model fall into four categories.
The first category, which represents
fundamental constants,
includes$\,^{1,2})$ \\
\begin{center}
\begin{tabular}{rcrll}
$M_{\rm H}$    & = &    7.289034  & \hm MeV             & hydrogen-atom mass
excess \\
$M_{\rm n}$    & = &    8.071431  & \hm MeV             & neutron mass excess\\
$e^2$          & = &    1.4399764 & \hm MeV fm \mbox{            }& electronic
charge squared\\
\end{tabular}\\[2ex]
\end{center}
One should note that for consistency we here use the same values for the
fundamental constants as in our 1981 mass calculation$\,^{1,2})$.
Results of a more recent evaluation of the fundamental constants appear
in Refs.$\,^{43,44})$.

The second category, which represents
constants that
have been determined from considerations other than nuclear masses,
includes$\,^{1-4})$ \\
\begin{center}
\begin{tabular}{rcrll}
$a_{\rm el}$   & = &   $1.433\times 10^{-5}$ & \hm MeV & electronic-binding
constant\\
$K$            & = &       240     & \hm MeV \mbox{            } &
                                           nuclear compressibility constant \\
$r_{\rm p}$    & = &    0.80       & \hm fm  & proton root-mean-square radius\\
$r_0$          & = &    1.16       & \hm fm  & nuclear-radius constant\\
$a$            & = &    0.68       & \hm fm  & range of Yukawa-plus-exponential
potential\\
$a_{\rm den}$  & = &   0.70        & \hm fm  & range of Yukawa function used to
\\
               &   &               & \hm     & \hspace{2em}
                                      generate nuclear charge distribution\\
\end{tabular}\\[2ex]
\end{center}

The third category, representing those constants whose
values were obtained from consideration of odd-even mass
differences$\,^{9,24,42})$
and other mass-like
quantities, are \\
\begin{center}
\begin{tabular}{rcrll}
$r_{\rm mac}$   & = &   4.80  & \hm MeV \mbox{            } & average
pairing-gap constant\\
$h$             & = &   6.6   & \hm MeV & neutron-proton interaction constant\\
$W$             & = &  30     & \hm MeV &  Wigner constant \\
$L$             & = &   0     & \hm MeV & density-symmetry constant\\
$a_3$           & = &   0     & \hm MeV & curvature-energy constant\\[2ex]
\end{tabular}\\[2ex]
\end{center}
It should be noted that the final calculated mass excess
is strictly independent of the value used for $r_{\rm mac}$. This constant
affects only the division of the mass excess between a macroscopic
part and the remaining microscopic correction.
We will therefore not include $r_{\rm mac}$ when we later
count the number of constants in our mass model.
It is the pairing constant
$r_{\rm mic}$ which enters the microscopic model that affects the
mass excess. It will be discussed below.

Since $\mu_{\rm th}=0$ in our case, Eqs.\ (\ref{deveq13}) and (\ref{deveq14})
can be solved with the experimental
data set of 1654 masses with $Z\geq 8$ and $N\geq 8$$\,^{45})$
and  28 fission-barrier heights
to determine
the remaining macroscopic constants and the error of our model.
Because it is now clear that the measurements of the masses
of $^{31-34}$Na that are listed in the 1989 midstream
evaluation of Audi$\,^{45})$ are in error,
we have made four revisions.
 For $^{31,32}$Na
we use early results of mass measurements at TOFI$\,^{46})$.
The final, slightly different values
appear in Ref.$\,^{47})$. For $^{33}$Na we use results
of new measurements at GANIL$\,^{48})$.
The data point for $^{34}$Na is excluded.

To present all the
macroscopic model constants together we list them here
but discuss their adjustment later.
These constants are

\begin{center}
\begin{tabular}{rcrll}
$       a_1 $   & = &  16.247 & \hm MeV \mbox{            } &  volume-energy
constant\\
$       a_2 $   & = &  23.92  & \hm MeV &  surface-energy constant\\
$       J   $   & = &  32.73  & \hm MeV &  symmetry-energy constant\\
$       Q   $   & = &  29.21  & \hm MeV &  effective surface-stiffness
constant\\
$       a_0 $   & = &   0.0   & \hm MeV &  $A^0$ constant \\
$ c_{\rm a} $   & = & 0.436   & \hm MeV &  charge-asymmetry constant \\
$       C   $   & = &  60     & \hm MeV &  pre-exponential compressibility-term
                                        constant \\
$    \gamma $   & = & 0.831   & \hm     &  exponential compressibility-term
                                        range constant\\[2ex]
\end{tabular}
\end{center}
The pairing constant
$r_{\rm mic}$ which enters the microscopic model
is also determined in a least-squares minimization
with the above 1654 masses, although
no barrier heights were included in its
determination.
Once the value of $r_{\rm mic}$ had been determined
the adjustment routines were
run again, this time with barriers included, to yield the final
 values of the constants listed above. The value of $r_{\rm mic}$
will be given in the section on microscopic constants.
The resulting error
in the FRDM is
$\sigma _{\rm th}$ = 0.669 MeV\@.

For completeness we  also specify the mass-energy conversion
factor used in the interim 1989 mass evaluation. In this evaluation
the relation between atomic mass units and energy is given
by$\,^{45})$
\beq
1\; \; {\rm u} = 931.5014\; \; {\rm MeV}
\eeq{conver}
Although a more recent value has been adopted$\,^{43,44,49})$,
it is the above value, consistent with the 1989 interim
mass evaluation$\,^{45})$,
that should be used if our calculated mass excesses in MeV are converted to
atomic mass units.

\subsection{Finite-range liquid-drop model}

In the present version of our model the
contribution to the atomic mass excess from the FRLDM
macroscopic energy is given by
\newpage \vspace{-0.4in}
\begin{eqnarray}
\lefteqn{E_{\rm mac}^{\rm FL}(Z,N,{\rm shape}) =} \nonumber \\[2ex]
& &\begin{array}{rclr}
& &\begin{displaystyle}  M_{\rm H}Z+M_{\rm n}N \end{displaystyle}
& \mbox{  \phantom{mass}mass excesses of $Z$ hydrogen atoms and $N$ neutrons}
\nonumber \\[2ex]
&-&\lefteqn{ \begin{displaystyle} a_{\rm v}\left(1 - \kappa _{\rm v}I^2
\right)A \end{displaystyle}}  & \mbox{  volume energy} \nonumber \\[2ex]
&+& \lefteqn{ \begin{displaystyle}
a_{\rm s}\left(1- \kappa _{\rm s}I^2\right)B_1
A^{2/3} \end{displaystyle}} & \mbox{  surface energy}
\nonumber \\[2ex]
&+&a_0A^0   & \mbox{$A^0$ energy} \nonumber \\[2ex]
&+&\begin{displaystyle} c_1\frac{Z^2}{A^{1/3}}B_3
\end{displaystyle} & \mbox{  Coulomb energy} \nonumber \\[2ex]
&-&\begin{displaystyle} c_4\frac{Z^{4/3}}{A^{1/3}}
\end{displaystyle}  & \mbox{  Coulomb exchange correction }\nonumber \\[2ex]
&+&\begin{displaystyle}f(k_{\rm f}r_{\rm p})\frac{Z^2}{A}
\end{displaystyle}  & \mbox{  proton form-factor correction
                               to the Coulomb energy} \nonumber \\[3ex]
&-&c_{\rm a}(N-Z)  & \mbox{  charge-asymmetry energy} \nonumber \\[2ex]
&+& \lefteqn{ W \left(|I|+ \left\{ \begin{array}{lll} 1/A \; \; , &
\mbox{$Z$ and $N$ odd and equal} \\
 0 \; \; ,& {\rm otherwise}
\end{array}  \right. \right) }
& \mbox{  Wigner energy} \nonumber \\[5ex]
&+ & \lefteqn{ \left\{ \begin{array}{ll}
+\begin{displaystyle}
\; \overline{\Delta} _{\rm p} + \overline{\Delta} _{\rm n} -
\delta _{\rm np}\; \; ,
\end{displaystyle}
& \mbox{$Z$ {\rm and} $N$ {\rm odd}} \\[2ex]
+\begin{displaystyle}
\; \overline{\Delta} _{\rm p}  \; \; ,
\end{displaystyle}
& \mbox{$Z$ {\rm odd and} $N$ {\rm even}} \\[2ex]
+\begin{displaystyle} \; \overline{\Delta} _{\rm n} \; \; ,
\end{displaystyle} &
\mbox{$Z$ {\rm even and}  $N$ {\rm odd}} \\[2ex]
+\begin{displaystyle}
\; 0 \; \; ,
\end{displaystyle} &
\mbox{$Z$ {\rm and} $N$ {\rm even}}
\end{array} \right. }  & \mbox{  average pairing energy} \nonumber \\[12ex]
&-&a_{\rm el}Z^{2.39}  & \mbox{  energy of bound electrons }
\end{array}\\
\eeqar{macenera}
This expression differs from the corresponding one used in our earlier
calculations$\,^{1,2})$ only in the form of the
average pairing energy appearing in the next-to-last term.
For the average neutron pairing gap
$\overline{\Delta} _{\rm n}$, average proton pairing gap
$\overline{\Delta} _{\rm p}$, and
 average neutron-proton interaction energy
$\delta _{\rm np}$
 we now
use$\,^{9,24,42})$
\beq
\overline{\Delta}_{\rm n}=\frac{r_{\rm mac}B_{\rm s}}{N^{1/3}}
\eeq{pairna}
\beq
\overline{\Delta}_{\rm p}=\frac{r_{\rm mac}B_{\rm s}}{Z^{1/3}}
\eeq{pairpa}
\beq
\delta _{\rm np} = \frac{h}{B_{\rm s}A^{2/3}}
\eeq{pairnpa}
The zero reference point for the pairing energy now corresponds to
even-even nuclei rather than to halfway between even-even and odd-odd
nuclei.

In the above expressions the
quantities $c_1$ and $c_4$ are defined in terms of the electronic
charge $e$ and the nuclear-radius constant $r_0$ by
\beqar
c_1&=&\begin{displaystyle}
\frac{3}{5}\frac{e^2}{r_0}
\end{displaystyle}\nonumber  \\[2ex]
c_4&=&\begin{displaystyle}
\frac{5}{4}{\left(\frac{3}{2\pi}\right)}^{2/3}c_1
\end{displaystyle}
\eeqar{ccoef}
The quantity $f$ appearing in the proton form-factor
correction to the Coulomb energy is given by
\beq
f(k_{\rm F}r_{\rm p})
=  -\frac{1}{8}\frac{{r_{\rm p}}^2e^2}{{r_0}^3}
\left[\frac{145}{48}-\frac{327}{2880}(k_{\rm F}r_{\rm p})^2
+\frac{1527}{1209600}(k_{\rm F}r_{\rm p})^4\right]
\eeq{formfac}
where the Fermi wave number is
\beq
k_{\rm F} = {\left( \frac{9\pi Z}{4A} \right)}^{1/3}\frac{1}{r_0}
\eeq{fermiwave}
The relative neutron excess $I$ is
\beq
I=\frac{N-Z}{N+Z}=\frac{N-Z}{A}
\eeq{neutex}

The relative surface energy $B_{\rm s}$, which is the ratio of the surface area
of the nucleus at
the actual shape to the surface area of the nucleus at the spherical
shape, is given by
\beq
B_{\rm s} =\frac{A^{-2/3}}{4\pi {r_0}^2}  \int _S dS
\eeq{surf}
The quantity $B_1$ is the relative generalized
surface or nuclear energy in a
model that accounts for the effect of the finite range of the nuclear
force. It is given by
\beq
B_1=\frac{A^{-2/3}}{8 \pi ^2 {r_0}^2a^4}
\int  \! \!\int_{V} \left( 2-\frac{|{\rm {\bf r}}-{\rm {\bf r}}'|}{a}\right)
\frac{e^{-|{\rm {\bf r}}-{\rm {\bf r}}'|/a}}
{|{\rm {\bf r}}-{\rm {\bf r}}'|/a} {d^3r}\,{d^3r'}
\eeq{nucena}
The relative Coulomb energy $B_3$ is given by
\beq
B_3 =\frac{15}{32 \pi ^2} \frac{A^{-5/3}}{{r_0}^5}
\int \! \! \int_{V} \frac{{d^3r}\,{d^3r'}}{|{\rm {\bf r}}-{\rm {\bf r}}'|}
\left[ 1- \left( 1 + \frac{1}{2}
\frac{|{\rm {\bf r}}-{\rm {\bf r}}'|}
{a_{\rm den}} \right)
e^{-|{\rm {\bf r}}-{\rm {\bf r}}'|/a_{\rm den}} \right]
\eeq{coulena}

For spherical shapes the quantities $B_1$ and $B_3$
can be evaluated analytically. With
\beq
x_0=\frac{r_0A^{1/3}}{a} \; \; \; \; {\rm and} \; \; \; \;
y_0=\frac{r_0A^{1/3}}{a_{\rm den}}
\eeq{xyz}
one obtains
\beqar
B_1^{(0)}& = & 1-\frac{3}{{x_0}^2}+ \left( 1+x_0\right) \left( 2+\frac{3}{x_0}
+\frac{3}{{x_0}^2} \right) e^{-2x_0}
\nonumber \\[2ex]
B_3^{(0)}& = & 1-\frac{5}{{y_0}^2}
\left[1-\frac{15}{8y_0}+\frac{21}{8{y_0}^3}
-\frac{3}{4}\left(1+\frac{9}{2y_0}+
\frac{7}{{y_0}^2}+\frac{7}{2{y_0}^3}
\right) e^{-2y_0} \right ] \eeqar{b3sph}

The expression $B_3$ for the relative Coulomb energy yields
the energy for an arbitrarily
shaped, homogeneously charged, diffuse-surface nucleus to all orders in
the diffuseness constant $a_{\rm den}$. The constants in front of
the integrals for $B_1$
and $B_3$ have been chosen so that $B_1$ and $B_3$ are 1 for a sphere in
the limit in which the range $a$ and diffuseness
$a_{\rm den}$ are zero, in analogy with the definition of the quantities
$B_{\rm s}$ and $B_{\rm C}$ in the standard liquid-drop
model.

\subsection{Values of FRLDM macroscopic-model constants}

The constants appearing in the expression for the finite-range liquid-drop
macroscopic model fall into four categories.
The first category, which represents
fundamental constants,
includes$\,^{1,2})$ \\
\begin{center}
\begin{tabular}{rcrlll}
$M_{\rm H}$   &=&   7.289034  & \hm MeV             & & hydrogen-atom mass
excess \\
$M_{\rm n}$   &=&   8.071431  & \hm MeV             & & neutron mass excess\\
$e^2$         &=&   1.4399764 & \hm MeV fm \mbox{            }& & electronic
charge squared\\
\end{tabular}\\[2ex]
\end{center}

The second category, which represents
constants that
have been determined from considerations other than nuclear masses,
includes$\,^{1,2})$ \\
\begin{center}
\begin{tabular}{rcrlll}
$a_{\rm el}$  &=&  $1.433\times 10^{-5}$ & \hm MeV & & electronic-binding
constant\\
$r_{\rm p}$   &=&   0.80 & \hm fm \mbox{            }  & & proton
root-mean-square radius\\
$r_0$         &=&   1.16 & \hm fm & & nuclear-radius constant\\
$a$           &=&   0.68 & \hm fm & & range of Yukawa-plus-exponential
potential\\
$a_{\rm den}$ &=&   0.70 & \hm fm & & range of Yukawa function used to \\
              & &        & \hm    & & \hspace{2em}
                                         generate nuclear charge distribution\\
\end{tabular}\\[2ex]
\end{center}

The third category, representing those constants whose
values were obtained from consideration of odd-even mass
differences$\,^{9,24,42})$ and other mass-like
quantities, are \\
\begin{center}
\begin{tabular}{rcrlll}
$r_{\rm mac}$   & = &   4.80  & \hm MeV \mbox{            } & & average
pairing-gap constant\\
$h$             & = &   6.6   & \hm MeV & & neutron-proton interaction
constant\\
$W$             & = &  30     & \hm MeV & & Wigner constant \\
\end{tabular}\\[2ex]
\end{center}
It should be noted that the final calculated mass excess
is strictly independent of the value used for $r_{\rm mac}$. This constant
affects only the division of the mass excess between the macroscopic
part and the remaining microscopic correction.
We therefore do not include $r_{\rm mac}$ when we later
count the number of constants in our mass model.
It is the pairing constant
$r_{\rm mic}$ which enters the microscopic model that affects the
mass excess. It will be discussed below.

Since $\mu_{\rm th}=0$ in our case,
Eqs.\ (\ref{deveq13}) and (\ref{deveq14})
can be solved with
the experimental
data set of 1654 masses with $Z\geq 8$ and $N\geq 8$$\,^{45})$
and  28 fission-barrier heights
to determine
the remaining macroscopic constants and the error of our model.
To present  all the
macroscopic model constants together we list them here
but discuss their adjustment later. \nolinebreak
These \nolinebreak constants \nolinebreak are
\begin{center}
\begin{tabular}{rcrlll}
$     a_{\rm v} $   & = &  16.00126 & \hm MeV \mbox{            } & &
volume-energy     constant\\
$\kappa_{\rm v} $   & = &   1.92240 & \hm MeV & & volume-asymmetry  constant\\
$     a_{\rm s} $   & = &  21.18466 & \hm MeV & & surface-energy    constant\\
$\kappa_{\rm s} $   & = &   2.345   & \hm MeV & & surface-asymmetry constant\\
$       a_0 $       & = &   2.615   & \hm MeV & & $A^0$ constant \\
$ c_{\rm a} $       & = &   0.10289 & \hm MeV & & charge-asymmetry constant
\\[2ex]
\end{tabular}
\end{center}
The resulting error in the FRLDM is
$\sigma _{\rm th} = 0.779$ MeV\@.

\subsection{Microscopic model}

The shell-plus-pairing correction
$E_{\rm s+p}(Z,N,{\rm shape})$ is
the sum of
the proton shell-plus-pairing correction
and the neutron shell-plus-pairing correction, namely
\beq
E_{\rm s+p}(Z,N,{\rm shape}) =
E_{\rm s+p}^{\rm prot}(Z,{\rm shape}) +
E_{\rm s+p}^{\rm neut}(N,{\rm shape})
\eeq{sumspp}
We give here the equations for the neutron
shell-plus-pairing correction. Completely analogous
expressions hold for
protons. We have
\beq
E_{\rm s+p  }^{\rm neut}(N,{\rm shape})=
E_{\rm shell}^{\rm neut}(N,{\rm shape})+
E_{\rm pair }^{\rm neut}(N,{\rm shape})
\eeq{emicr}
Both terms are evaluated from a set of calculated single-particle
levels. As before, the shell correction is calculated
by use of Strutinsky's
method$\,^{25,26})$. Thus
\beq
E_{\rm shell}^{\rm neut}(N,{\rm shape})=
\sum_{i=1}^{N} e_i
-{\widetilde{E}}^{\rm neut}(N,{\rm shape})
\eeq{strut}
where $e_i$ are calculated single-particle energies
and ${\widetilde{E}}^{\rm neut}(N,{\rm shape})$ is
the smooth single-particle energy sum calculated in
the Strutinsky method.
The pairing correction is the difference between the pairing correlation
energy and the average pairing correlation energy, namely
\beq
E_{\rm pair}^{\rm neut}(N,{\rm shape}) =
 E_{\rm p.c.}^{\rm neut}(N,{\rm shape}) -
{\widetilde{E}}_{\rm p.c.}^{\rm neut}(N,{\rm shape})
\eeq{paircor}
where  $E_{\rm p.c.}^{\rm neut}(N,{\rm shape})$ is given
by Eq.~(\ref{epairln}) below and
${\widetilde{E}}_{\rm p.c.}^{\rm neut}(N,{\rm shape})$
is given by Eq.~(\ref{totavpln})
below.
For the pairing correction we now use the
Lipkin-Nogami$\,^{21-23})$
version of the BCS method,
which takes into account the lowest-order correction to the total
energy of the system
associated with
particle-number fluctuation.

The single-particle potential felt by a nucleon is given by
\beq
V=V_1+V_{\rm s.o.}+V_{\rm C}
\eeq{singpot}
The first term is the spin-independent nuclear part of the potential,
which is calculated in terms of the folded-Yukawa potential
\beq
V_1({\rm {\bf r}})=  -\frac{V_0}{4\pi {a_{\rm pot}}^3}
\int_{\rm V}\frac{e^{-|{\rm {\bf r}}-{\rm {\bf r}}'|/a_{\rm pot}}}
{|{\rm {\bf r}}-{\rm {\bf r}}'|/a_{\rm pot}} {d^3r'}
\eeq{yukpot}
where the integration is over the volume of the generating shape,
whose volume is held fixed at $\frac{4}{3}\pi {R_{\rm pot}}^3$ as the shape
is deformed. The potential radius $R_{\rm pot}$ is given by
\beq
R_{\rm pot} = R_{\rm den} + A_{\rm den} - B_{\rm den}/R_{\rm den}
\eeq{rpot}
with
\beq
R_{\rm den}= r_0 A^{1/3}(1+\overline{\epsilon})
\eeq{rden}
The potential depth
$V_{\rm p}$ for protons and potential depth $V_{\rm n}$ for
neutrons are given by
\beq
V_{\rm p} = V_{\rm s} + V_{\rm a}\overline{\delta}
\eeq{vp}
\beq
V_{\rm n} = V_{\rm s} - V_{\rm a}\overline{\delta}
\eeq{vn}

The average bulk nuclear asymmetry
$\overline{\delta}$
appearing in Eqs.~(\ref{vp}) and (\ref{vn})
and average relative deviation $\bar{\epsilon}$ in the
bulk of the density appearing in Eq.~(\ref{rden}) are
given by the droplet model and thus depend on
the values of  the droplet-model constants. The
FRDM macroscopic constants are determined in a nonlinear
least-squares adjustment, which requires about  1000 steps
to find the optimum constants. In principle, these constants
should then be used in
the determination of the single-particle potential, the
potential-energy surfaces should be recalculated with the
new constants, a new mass calculation should be performed, and
a new set of macroscopic constants should be determined, with
this iteration repeated until convergence. Because the
calculation of potential-energy surfaces is extremely time-consuming,
only one iteration has been performed.

Furthermore, in determining the single-particle potential we
have used the following early forms$\,^{50})$ of the
droplet model expressions for $\bar{\delta}$ and $\bar{\epsilon}$:
\beq
\overline{\delta} =
\left(I+\frac{3}{8}\frac{c_1}{Q}\frac{Z^2}{A^{5/3}}\right)/
\left(1+\frac{9}{4}\frac{J}{Q}\frac{1}{A^{1/3}}\right)
\\[2ex] \eeq{name8old}
\beq
\overline{\epsilon} = \left( -\frac{2a_2}{A^{1/3}}
+L\overline{\delta} ^2 + c_1\frac{Z^2}{A^{4/3}}\right)/K \\[2ex]
\eeq{name6old}
The range
\beq
a_{\rm pot}=0.8 \; {\rm fm}
\eeq{apot}
of the Yukawa function in Eq.~(\ref{yukpot}) has been determined from
an adjustment of calculated single-particle levels to experimental data
in the rare-earth and actinide
regions$\,^{34})$. It is kept constant for
nuclei throughout the periodic system.

The spin-orbit potential is given by the expression
\beq
V_{\rm s.o.}=  -\lambda {\left( \frac{\hbar}{2m_{\rm nuc}c} \right)}^2
\frac{ \mbox{\boldmath $\sigma$} \cdot
\nabla V_1 \times \mbox{\boldmath $p$}}{\hbar}
\eeq{spinor}
where $\lambda$ is the spin-orbit interaction strength,
$m_{\rm nuc}$ is the nucleon mass,
{\boldmath $\sigma$} is the Pauli spin matrix, and
{\boldmath $p$} is the nucleon momentum.

The spin-orbit strength has been determined from adjustments to
experimental levels in the rare-earth and actinide regions. It has
been shown$\,^{1,14,34})$ that many nuclear
properties throughout the periodic system are well reproduced
with $\lambda$ given by a function linear in $A$ through the values
determined in these two regions. This gives
\beq
\lambda_{\rm p}= 6.0\left( \frac{A}{240} \right) + 28.0
= 0.025A+ 28.0 = k_{\rm p}A + l_{\rm p}
\eeq{lamp}
for protons and
\beq
\lambda_{\rm n}= 4.5\left( \frac{A}{240} \right) + 31.5
= 0.01875A+ 31.5 = k_{\rm n}A + l_{\rm n}
\eeq{lamn}
for neutrons.

Finally, the Coulomb potential for protons is given by
\beq
V_{\rm C}({\rm {\bf r}})=e\rho_{\rm c}
\int_{\rm V}\frac{d^3r'}
{|{\rm {\bf r}}-{\rm {\bf r}}'|}
\eeq{coulpot}
where the charge density $\rho _{\rm c}$ is given by
\beq
\rho_{\rm c}=\frac{Ze}{\frac{4}{3}\pi A{r_0}^3}
\eeq{chargeden}

The number of basis functions used in our calculations is
\beq
N_{\rm bas} = 12
\eeq{nbas}
The overall curvature of the basis functions is chosen to
yield
\beq
\hbar \omega_0 = C_{\rm cur}/A^{1/3}
\eeq{hbarom}
with
\beq
C_{\rm cur} = 41\; \; {\rm MeV}
\eeq{hbaromcon}

\subsection{Microscopic pairing models}

Because of its basic simplicity, the BCS pairing
model$\,^{51-54})$
has been the pairing model of choice in
most previous nuclear-structure
calculations$\,^{1,2,29,55})$.
However, a well-known deficiency
of the BCS model is that for large spacings between
the single-particle levels at the Fermi surface,
no non-trivial solutions exist.
In practical applications, these situations occur not only
at magic numbers, but also, for example, for deformed actinide nuclei
at neutron numbers $N=142$ and
152. By taking into account
effects associated with particle-number fluctuations,
the Lipkin-Nogami approximation$\,^{21-23})$
goes beyond the BCS approximation and
avoids such collapses.

In solving the pairing equations for
neutrons or protons in either the BCS or Lipkin-Nogami model,
we consider a constant pairing interaction
$G$ acting between \mbox{$N_2 - N_1 + 1$} doubly degenerate
single-particle levels,
which are occupied by $N_{\rm int}$ nucleons.
This interaction interval starts at level $N_1$, located below the
Fermi surface, and ends at level $N_2$, located above the Fermi surface.
With the definitions we use here, the levels are numbered consecutively
starting with number 1 for the level at the bottom of the well.
Thus, for even particle numbers,
the last occupied levels in the neutron and proton wells
are $N/2$ and $Z/2$, respectively.

The  level pairs included in the
pairing calculation are
often chosen symmetrically around the Fermi surface.
However, for spherical nuclei it is more reasonable to
require that degenerate spherical states have equal occupation probability.
This condition cannot generally be satisfied simultaneously
with a symmetric choice of levels in the interaction region.
We therefore derive the pairing equations below for the more general case
of arbitrary  $N_1$ and $N_2$.

In the Lipkin-Nogami pairing model$\,^{21-23})$
the pairing gap $\Delta$, Fermi energy $\lambda$,
number-fluctuation constant $\lambda_2$,  occupation probabilities
${v_k}^2$, and shifted single-particle energies $\epsilon_k$
are determined from the  $2(N_2 - N_1) + 5$
coupled nonlinear equations
\beq
 N_{\rm tot} = 2 \sum_{k=N_1}^{N_2} {v_k}^2 + 2(N_1 - 1)
\eeq{numbercon}
\beq
\frac{2}{G} = \sum_{k=N_1}^{N_2} \frac{1}
{\sqrt{(\epsilon_k - \lambda)^2 + {\Delta}^2}}
\eeq{gap}
\beq
{v_k}^2 = \frac{1}{2}\left[1-\frac{\epsilon_k-\lambda}
{\sqrt{(\epsilon_k - \lambda)^2 + {\Delta}^2}}\right],\; \; \;
k=N_1,N_1+1,\ldots,N_2\\[1ex]
\eeq{occpro}
\beq
\epsilon_k = e_k +(4\lambda_2-G){v_k}^2, \; \; \; k=N_1,N_1+1,\ldots,N_2\\[1ex]
\eeq{epsk}
\beq
\lambda_2 = \frac{G}{4}\left[\frac{
\begin{displaystyle}
\left( \sum_{k=N_1}^{N_2} {u_k}^3{v_k}  \right)
\left( \sum_{k=N_1}^{N_2} {u_k}  {v_k}^3\right) -
       \sum_{k=N_1}^{N_2} {u_k}^4{v_k}^4
\end{displaystyle}
}
{\begin{displaystyle}
\left( \sum_{k=N_1}^{N_2} {u_k}^2{v_k}^2\right)^2 -
       \sum_{k=N_1}^{N_2} {u_k}^4{v_k}^4
\end{displaystyle}
}
\right]
\eeq{lambda2}
where
\beq
{u_k}^2 = 1 -{v_k}^2 \; \; , \; \; k=N_1,N_1+1,\ldots,N_2
\eeq{uk}

The quasi-particle energies $E_k$ of the odd nucleon
in an odd-$A$ nucleus are now given by$\,^{22})$
\beq
E_k= \left[ (\epsilon_k-\lambda)^2 + \Delta^2 \right]^{1/2}
+\lambda_2,\; \; \;
k=N_1,N_1+1,\ldots,N_2
\eeq{quasilip1}
In the Lipkin-Nogami model it is the sum $\Delta + \lambda_2$
that is identified with odd-even mass differ\-ences$\,^{22})$.
We denote this sum by $\Delta_{\rm LN}$.

The pairing-correlation energy plus quasi-particle
energy  in the Lipkin-Nogami model is given by
\beq
E_{\rm p.c.} = \sum_{k=N_1}^{N_2}(2{v_k}^2-n_k)e_k
-\frac{\Delta^2}{G} -
\frac{G}{2}\sum_{k=N_1}^{N_2}(2{v_k}^4-n_k)
-4\lambda_2\sum_{k=N_1}^{N_2}{u_k}^2{v_k}^2 +E_i\theta_{{\rm odd},N_{\rm tot}}
\eeq{epairln}
where $e_k$ are the single-particle energies and $n_k$, with
values 2, 1, or 0, specify the sharp distribution of particles in the
absence of pairing.
The quasi-particle energy $E_i$ for the odd particle
occupying level $i$ is given by Eq.~(\ref{quasilip1}),
and $\theta_{{\rm odd},N_{\rm tot}}$ is unity if $N_{\rm tot}$
is odd and zero if $N_{\rm tot}$ is even.

\subsection{Effective-interaction pairing-gap models}

In microscopic pairing calculations the pairing strength
$G$ for neutrons and protons can be obtained from effective-interaction
pairing gaps $\Delta_{G_{\rm n}}$ and
$\Delta_{G_{\rm p}}$ given by$\,^{9})$

\beq
\Delta_{G_{\rm n}}=\frac{r_{\rm mic}B_{\rm s}}{N^{1/3}}
\eeq{pairnmic}
\beq
\Delta_{G_{\rm p}}=\frac{r_{\rm mic}B_{\rm s}}{Z^{1/3}}
\eeq{pairpmic}
The dependence of the pairing strength $G$ on the
corresponding effective-interaction pairing gap
${\Delta_G}$
is obtained from the
microscopic equations
by assuming a constant level
density for the average nucleus in the vicinity of the Fermi surface.
This allows the sums in the equations to be replaced by integrals.
The average level density of doubly degenerate levels is taken to be
\beq
\widetilde{\rho} = \frac{1}{2} \widetilde{g}(\widetilde{\lambda})
\eeq{avden}
where $\widetilde{g}$ is the smooth level density that is obtained
in  Strutinsky's shell-correction method and $\widetilde{\lambda}$ is
the Fermi energy of the smoothed
single-particle energy$\,^{29,56})$.
Thus, we can make the substitution
\beq
\sum_{k=N_1}^{N_2}f(e_k - \lambda)
\Longrightarrow \widetilde{\rho}\int_{y_1}^{y_2}f(x)dx
\eeq{subst}
where
\beqar
y_1& = &\frac{ -
\frac{1}{2}N_{\rm tot} +N_1 - 1
}{\widetilde{\rho}} \nonumber \\[1ex]
y_2& = &\frac{ -
\frac{1}{2}N_{\rm tot} + N_2
}{\widetilde{\rho}}
\eeqar{limits}

The gap equation~(\ref{gap}) can now be evaluated for an {\it average}
nucleus, with the result
\beqar
\frac{1}{G} &=& \frac{1}{2}\widetilde{\rho} \int^{y_2}_{y_1}
\frac{dx}{\sqrt{x^2 + {\Delta_G}^2}} \nonumber \\[1ex]
 & = & \frac{1}{2} \widetilde{\rho}\left[ \ln \!
       \left(\sqrt{{y_2}^2 +{\Delta_G}^2}
+y_2 \right) - \ln \! \left(\sqrt{{y_1}^2+{\Delta_G}^2} + y_1\right) \right]
\eeqar{avgap}
{}From this expression, the pairing strength
$G$ in the BCS model can be determined in any region of the nuclear chart.

The same expression may also be used in the
Lipkin-Nogami case, but some reinterpretations
are necessary. It is now the energies
$\epsilon_k$ occurring in Eq.~(\ref{gap})
that are assumed to be equally spaced.
These are not precisely the single-particle
energies $e_k$ but are related to them  by Eq.~(\ref{epsk}). Thus, in order
for $\epsilon_k$ to be equally spaced, the single-particle  energies $e_k$ must
be shifted downward by the amounts $(4\lambda_2 - G){v_k}^2$.
Since the occupation probability ${v_k}^2$ is approximately unity far below
the Fermi surface and zero far above, the corresponding single-particle energy
distribution is approximately uniform far above and far below the Fermi surface
but spread apart by the additional amount $4\lambda_2 - G$
close to the Fermi surface.
Although this decrease in level density near the Fermi surface is accidental,
it is in approximate accord with the ground-state structure of real
nuclei, since
the increased stability associated with ground-state
configurations is due to low level
densities near the Fermi surface$\,^{24,56})$.

In the Lipkin-Nogami model, it is the quantity $\Delta +\lambda_2$ that is
associated with odd-even mass differences,
whereas in the BCS model it is $\Delta$ only that should be directly compared
to the experimental data. This leads to the expectation that there is
a related difference between $\Delta^{\rm LN}_G$ and $\Delta_G^{\rm BCS}$,
the effective-interaction pairing gaps associated with the LN and
BCS models, respectively.
Since we determine the constants of the model for $\Delta_G^{\rm LN}$
directly from least-squares minimization, it is not necessary
to specify exactly such a relationship.
However, the above
observation is of value as a rough rule of thumb, and
to remind us to expect that the effective-interaction pairing gaps in
the BCS and LN models are of somewhat different magnitude.

The expression for the {\it average}
pairing correlation  energy plus quasi-particle energy
$\widetilde{E}_{\rm p.c.}$ in  the Lipkin-Nogami model
is obtained in a similar manner as the
expression for the pairing matrix element $G$.
For the average pairing correlation
energy  plus quasi-particle energy  in the Lipkin-Nogami model we then obtain
\beqar
\widetilde{E}_{\rm p.c.}
& = &  \frac{1}{2}\widetilde{\rho}\left[(y_2-G)
\left(y_2-\sqrt{{y_2}^2+{{\Delta_G}}^2} \right)
+(y_1 - G)
\left(y_1+\sqrt{{y_1}^2+{{\Delta_G}}^2}\right) \right]
 \nonumber \\[2ex]
&  &  \mbox{}+\frac{1}{4}(G-4\widetilde{\lambda}_{2})\widetilde{\rho}
{{\Delta_G}}
\left[ \tan^{-1} \! \left(\frac{y_2}{{{\Delta_G}}}\right)
 - \tan^{-1} \! \left(\frac{y_1}{{{\Delta_G}}}\right)\right]
+ \overline{\Delta}\theta_{{\rm odd},N_{\rm tot}}
\eeqar{totavpln}

The expression for $\widetilde{\lambda}_{2}$ for an
average nucleus is fairly lengthy.
It is given by
\beqar
\widetilde{\lambda}_{2} = \frac{G}{4}
\left( \frac{A - C}{B - C} \right)
\eeqar{clambda2}
where
\beqar
A & = &
\left(\frac{\widetilde{\rho}{\Delta_G}}{4}\right)^2 \left\{
\left(\frac{2}{G\widetilde{\rho}}
\right)^2
- \left[ \ln \!  \left(
\frac{\sqrt{{y_2}^2+{{\Delta_G}}^2}}
     {\sqrt{{y_1}^2+{{\Delta_G}}^2}}
\right)
\right]^2
\right\}
\nonumber \\[3ex]
B & = &
\frac{{{\Delta_G}}^2{\widetilde{\rho}}^2}{16}\left[
\tan^{-1} \! \left(\frac{y_2}{{{\Delta_G}}}\right)-
\tan^{-1} \! \left(\frac{y_1}{{{\Delta_G}}}\right)\right]^2
\nonumber \\[3ex]
C & = &
\frac{\widetilde{\rho}{\Delta_G}}{32}
\left[{\Delta_G}\left(
  \frac{y_2}{{y_2}^2 +{\Delta_G}^2}
 -\frac{y_1}{{y_1}^2 +{\Delta_G}^2} \right)
+ \tan^{-1} \! \left( \frac{y_2}{{\Delta_G}}\right)
- \tan^{-1} \! \left( \frac{y_1}{{\Delta_G}}\right)
\right]
\eeqar{cclambda2}

\subsection{Shell correction}

The Strutinsky shell-correction method$\,^{25,26})$
requires two additional constants, the order $p$ and
the range $\gamma$. The shell correction
should be insensitive to these quantities
within a certain range of values.
Their values can therefore be determined in principle by
requiring  the plateau condition to be fulfilled.
We have found that for heavy nuclei
this condition is indeed fulfilled, with the shell correction
for nuclear
ground-state shapes
insensitive to the values of these two constants. However, for light nuclei
this is no longer the
case. Here the shell correction may vary by several MeV for a reasonable
range of values of the range $\gamma$. Moreover, the shell correction often
does not exhibit any plateau. This probably indicates a gradual
breakdown of the shell-correction method as
one approaches the very lightest region
of nuclei, where the number of
single-particle levels is small.

In the present calculation we choose
\beq
p=8
\eeq{strutord}
for the order in the Strutinsky shell-correction method.
The corresponding range $\gamma$ is given by
\beqar
\gamma = C_{\rm sr}\hbar \omega_0 B_{\rm s}
\eeqar{strutran}
with
\beq
C_{\rm sr}= 1.0
\eeq{strutranv}
and $B_{\rm s}$ given by Eq.~(\ref{surf}).
This choice lowers the error of the mass model
to 0.669 MeV from 0.734 MeV obtained with
the same range coefficient but no
dependence on surface area in a sixth-order correction.

\subsection{Values of microscopic-model constants}

The constants appearing in the expressions occurring in the
microscopic shell-plus-pairing calculation  fall into four categories.
The first category, which represents
fundamental constants,
includes
\begin{center}
\begin{tabular}{rcrll}
$m_{\rm nuc}$ & = & 938.90595  & \hm MeV    &   nucleon mass \\
$\hbar c$     & = & 197.32891  & \hm MeV fm \mbox{            } &   Planck's
constant multiplied\\
              &   &            & \hm        &  \hspace{2em} by the speed of
light and
                                                 divided by $2\pi$\\
$e^2$         & = & 1.4399764  & \hm MeV fm &   electronic charge squared
\end{tabular}
\end{center}
The
electronic charge squared has already been counted
among the macroscopic constants.

The second category, which represents
constants that
have been determined from considerations other than nuclear masses,
includes$\,^{1,2,29})$ \\
\begin{center}
\begin{tabular}{rcrll}
$C_{\rm cur}$    & = & 41      & \hm MeV \mbox{            } &  basis curvature
constant \\
$  V_{\rm s} $   & = & 52.5    & \hm MeV & symmetric  potential depth
constant\\
$  V_{\rm a} $   & = & 48.7    & \hm MeV & asymmetric potential depth constant
\\
$  A_{\rm den} $ & = & 0.82    & \hm fm  & potential radius correction
constant\\
$  B_{\rm den} $ & = & 0.56    & \hm ${\rm fm}^2$ & potential radius
curvature-correction
                                        constant \\
$  a_{\rm pot} $ & = & 0.8     & \hm fm  & potential diffuseness constant\\
$  k_{\rm p}   $ & = & 0.025   & \hm     & proton spin-orbit $A$ coefficient \\
$  l_{\rm p}   $ & = & 28.0    & \hm     & proton spin-orbit constant \\
$  k_{\rm n}   $ & = & 0.01875 & \hm     & neutron spin-orbit $A$ coefficient
\\
$  l_{\rm n}   $ & = & 31.5    & \hm     & neutron spin-orbit constant \\
\end{tabular}
\end{center}

The third category, representing those constants whose
values were obtained from consideration of  mass-like
quantities, are \\
\begin{center}
\begin{tabular}{rcrll}
$N_{\rm bas}$     & = &  12 &  \mbox{            } &number of basis functions
\\
$ p$              & = &   8 &             &  order of Strutinsky shell
correction \\
$C_{\rm sr}$      & = & 1.0 &             &  Strutinsky range coefficient \\
\end{tabular}
\end{center}

The fourth category, representing those constants whose
values were obtained from  a least-squares adjustment simultaneously
with the macroscopic constants of the FRDM\@, includes
only one microscopic constant, namely
\begin{center}
\begin{tabular}{rcrll}
$r_{\rm mic}$ & = & 3.2 & \hm MeV \mbox{            }&
                             LN effective-interaction pairing-gap constant
\end{tabular}
\end{center}

In addition, the following
droplet-model constants,
which have been determined in an earlier study$\,^{50})$,
are used in the
expressions for
the $average$ bulk nuclear asymmetry $\overline{\delta}$ and
{\it  average\/} relative deviation $\overline{\epsilon}$
in the bulk density that
are used to calculate $V_{\rm p}$, $V_{\rm n}$,
and $R_{\rm den}$
in Eqs.~(\ref{vp}), (\ref{vn}), and (\ref{rden}), respectively:
\begin{center}
\begin{tabular}{rcrll}
$a_1$ &= & 15.677    & \hm MeV \mbox{            }& volume-energy constant\\
$a_2$ &= & 22.00     & \hm MeV \mbox{            }& surface-energy constant\\
$J $  &= & 35        & \hm MeV              & symmetry-energy constant\\
$L$   &= &  99       & \hm MeV              & density-symmetry constant\\
$Q $  &= & 25        & \hm MeV              & effective surface-stiffness
constant   \\
$K$   &= & 300       & \hm MeV              & compressibility constant \\
$r_0$ &= & 1.16      & \hm fm               & nuclear-radius constant\\
\end{tabular}\\[2ex]
\end{center}
Use of these values in Eqs.~(\ref{name8old}) and (\ref{name6old})
leads to
\beq
\overline{\delta}=
\begin{displaystyle}
\frac{(N-Z)/A+0.0112Z^2/A^{5/3}}
{1+3.15/A^{1/3}}
\end{displaystyle}
\eeq{delbars}
\beq
\overline{\epsilon}=  -\frac{0.147}{A^{1/3}}
+0.330{\overline{\delta}}^2 + \frac{0.00248Z^2}{A^{4/3}}
\eeq{epsbars}
One could in principle carry through
the iterations discussed above to obtain a consistent set of
droplet-model constants for the macroscopic part and for the
single-particle potential, but the
required computational
effort would be extensive.
However, the value of $r_0$ is precisely the same as that
used in the macroscopic model.

\section{Enumeration of constants}

It is always of interest to have a clear picture of exactly what
constants enter a model. Naturally, anyone who sets out to
verify a calculation by others or uses a model for new
applications needs
a complete  specification of the model, for which
a full specification of the constants and their values is an
essential part. Also, when different models are compared it is
highly valuable to fully understand exactly what constants
enter the models. Unfortunately, discussions of model
constants are often incomplete, misleading, and/or erroneous.
For example, in Table A of Ref.$\,^{57})$ the
number of parameters of the mass model of Spanier and
Johansson$\,^{58})$ is
listed as 12. However, in the article$\,^{58})$ by Spanier and Johansson
the authors themselves list in their Table A
30 parameters plus 5 magic numbers that are not calculated
within the mass model and must therefore be considered parameters, for
a total of  at least 35 parameters.

We  specify here {\it all} the constants that
enter our model, rather than just those  that
in the final step are adjusted to experimental data by
a least-squares procedure. We
also include
such constants as the number of basis functions used
and  fundamental constants like the electronic charge and Planck's constant.

\subsection{Constants in the FRDM}

The discussion in the previous section allows us to enumerate the
constants in the FRDM model in Table~\ref{frdmp}.
\begin{table}[t]
\begin{small}
\begin{center}
\caption[frdmp]{{Constants in the FRDM\@.} {\baselineskip=12pt\small The
third column \label{frdmp} gives
the  number of constants
adjusted to nuclear masses or mass-like quantities such as
odd-even mass
differences or fission-barrier heights.
The fourth column gives the number of constants
determined from other considerations.}\\}
\begin{tabular}{llrr}
  \hline\\[-0.07in]
Constants & Comment & Mass-like & Other \\[0.08in]
  \hline\\[-0.07in]
$M_{\rm H}$, $M_{\rm n}$, $e^2$
& Macroscopic fundamental constants                           & 0 & 3
\\[0.08in]
$a_{\rm el}$, $r_0$, $r_{\rm p}$, &
Macroscopic constants from considerations                     & 0 & 6 \\
$a$, $a_{\rm den}$, $K$
& other than mass-like data                                   &   &
\\[0.08in]
$L$, $a_3$, $W$, $h$ &  Macroscopic constants obtained
                                                              &  4 &  0\\
 & in prior adjustments to mass-like data                      & & \\[0.08in]
$a_1$, $a_2$, $J$, $Q$, $a_0$, &
Macroscopic constants determined by                           & 8 &  0 \\
 $C$, $\gamma$, $c_{\rm a}$&
current least-squares adjustments                             &   &
\\[0.08in]
$\hbar c$, $m_{\rm nuc}$ &
 Microscopic fundamental constants                             & 0 & 2
\\[0.08in]
$V_{\rm s}$, $V_{\rm a}$, $A_{\rm den}$, $B_{\rm den}$,
$C_{\rm cur}$, & Microscopic constants \hfill                 & 0 & 10\\
$k_{\rm p}$, $l_{\rm p}$, $k_{\rm n}$, $l_{\rm n}$,
$a_{\rm pot}$
                                                              &   &  \\[0.08in]
 $N_{\rm bas}$, $p$,
$C_{\rm sr}$
& Microscopic constants determined                           & 3  & 0  \\
& from considerations of mass-like quantities                  & &  \\[0.08in]
  $r_{\rm mic}$
& Microscopic constant determined by                       & 1 & 0 \\
& current least-squares adjustments                         & &  \\[0.08in]
$a_1$, $a_2$, $J$, $K$, $L$, $Q$ &
Droplet-model constants that enter the single-
                          & 0 & 0\\
 & particle potential (see discussion in text)                  & & \\[0.08in]
  \hline\\[-0.07in]
\multicolumn{2}{l}{Subtotals}   & 16 & 21 \\[0.08in]
  \hline\\[-0.07in]
\multicolumn{2}{l}{Total}          & & 37 \\[0.08in]
\hline
\end{tabular}
\end{center}
\end{small}
\end{table}
{}From this list  we see that the macroscopic-microscopic method
requires relatively few constants.
One feature of the model gives rise to a small complication
when counting the number of constants. Droplet-model constants
occur also in the determination
 of the single-particle
potential.
\begin{table}[t]
\begin{small}
\begin{center}
\caption[frldmp]{{Constants in the FRLDM\@.} {\baselineskip=12pt\small The
third column \label{frldmp} gives
 the  number of constants
adjusted to nuclear masses or mass-like quantities such as
odd-even mass
differences or fission-barrier heights.
The fourth column gives the number of constants
determined from other considerations.}\\}
\begin{tabular}{llrr}
  \hline\\[-0.07in]
Constants & Comment & Mass-like & Other \\[0.08in]
$M_{\rm H}$, $M_{\rm n}$, $e^2$
& Macroscopic fundamental constants                           & 0 & 3
\\[0.08in]
$a_{\rm el}$, $r_0$, $r_{\rm p}$, &
Macroscopic constants from considerations                     & 0 & 5 \\
$a$, $a_{\rm den}$
& other than mass-like data                                   &   &
\\[0.08in]
 $W$, $h$ &  Macroscopic constants obtained
                                                              &  2 &  0\\
 & in prior adjustments to mass-like data                       & & \\[0.08in]
$a_{\rm v}$, $\kappa_{\rm v}$, $a_{\rm s}$, $\kappa_{\rm s}$,   &
Macroscopic constants determined by                           & 6  &  0 \\
 $a_0$, $c_{\rm a}$&
current least-squares adjustments                            &   &  \\[0.08in]
$\hbar c$, $m_{\rm nuc}$ &
 Microscopic fundamental constants                            & 0 & 2\\[0.08in]
$V_{\rm s}$, $V_{\rm a}$, $A_{\rm den}$, $B_{\rm den}$,
$C_{\rm cur}$, & Microscopic constants \hfill                & 0 & 10\\
$k_{\rm p}$, $l_{\rm p}$, $k_{\rm n}$, $l_{\rm n}$,
$a_{\rm pot}$
&
                                                             &   &  \\[0.08in]
 $N_{\rm bas}$, $p$, $C_{\rm sr}$, $r_{\rm mic}$  & Microscopic
constants determined                                         & 4 & 0 \\
& from considerations of mass-like quantities \hfill             & &
\\[0.08in]
$a_1$, $a_2$, $J$, $K$, $L$, $Q$ &
Droplet-model constants that enter the single-
                          & 3 & 0\\
 & particle potential
 (see discussion in text)                                      & & \\[0.08in]
  \hline\\[-0.07in]
\multicolumn{2}{l}{Subtotals}   & 15 & 20 \\[0.08in]
  \hline\\[-0.07in]
\multicolumn{2}{l}{Total} & &          35 \\[0.08in]
\hline
\end{tabular}
\end{center}
\end{small}
\end{table}
However, a different set of constants is used here because,
as discussed above, one does not know
what the optimum values are until the calculation
has been completed.
In principle, the calculation should
be repeated with the new  droplet-model constants
defining the single-particle potential
until  convergence is obtained. In Table~\ref{frdmp}  we have
counted the number of constants as if this procedure had been carried out.

However, since the droplet-model constants used in
the present
calculations are different in the microscopic part and in the macroscopic
part, different counting
schemes could also be employed. Since
the droplet-model constants used in the microscopic expressions are
obtained from four primary constants$\,^{50})$ and
nuclear masses were used
only to give rough estimates of these constants,
one may not wish to regard them
as determined from
mass-like quantities.  One of the four
primary constants is
the nuclear radius constant $r_0$, which has the same value as we use in our
macroscopic model. Therefore, only three remain that could
be considered as additional FRDM  constants.
With this classification scheme
the number of constants adjusted to mass-like quantities remains
16 and the total number of constants in the model increases from
37 to 40. Alternatively, if we do count the three primary constants
as adjusted to nuclear masses, the total number of FRDM
constants  is 40, while the number adjusted to mass-like
quantities increases from 16 to 19.

\subsection{Constants in  the FRLDM}
The constants in the FRLDM, which are either
identical to or similar to the constants in the FRDM,
are enumerated in Table~\ref{frldmp}.
We mentioned in the discussion of the FRDM constants
that the six constants in the last line of Table~\ref{frdmp} would
converge to the values of the same constants listed earlier
in the table after a sufficient number of iterations.
In the FRDM these constants therefore need not be regarded as
additional constants. In contrast, in the FRLDM they must be regarded as
constants obtained from adjustments to mass-like
quantities. However, as mentioned in the discussion of
the FRDM constants, these constants are all obtained from
three primary constants, so we only include three
in this category.

\section{Results}

\subsection{Determination of ground-state shapes \label{4.1} and masses}

The adjustment of constants
in  the macroscopic model is simplified enormously because the ground-state
shape and fission saddle-point shape are approximately independent of the
precise values of these constants when they are
varied within a reasonable range$\,^{59})$. We therefore
calculate the ground-state deformation with one set of constants
and subsequently determine the various terms in the mass expression
at this deformation. The constants of the macroscopic model can then
be adjusted, with the nuclear
shapes remaining fixed.

A significant advantage of this approach is that the effect of new features
can often be investigated without repeating the entire calculation
from the beginning, which would take about 100 hours of CRAY-1 CPU time.
For example, when we investigated different pairing models and determined
the optimum
value of the pairing constant, we needed to recalculate only the
pairing-energy term for each of the 8979 nuclei in our study. Since
we have in the initial part of the calculation determined ground-state
shapes and stored the corresponding ground-state single-particle levels
for all nuclei on disc, we need only read in the single-particle levels,
do the pairing calculation, and readjust the model constants to
obtain the effect of a new pairing model or new pairing-model constant.
Such a study takes only about 20 minutes of CRAY-1 CPU time.

Our determination of mass-model constants and ground-state
nuclear masses involves several steps. We first briefly list
these steps and then continue with a more extensive discussion.

\begin{enumerate}

\item
Potential-energy surfaces are calculated versus $\epsilon_2$ and
$\epsilon_4$. In this calculation, which was actually performed already
in 1987, the FRLDM as defined in Ref.$\,^{3})$ is used, except
that for the pairing calculations the BCS approximation is used instead
of the LN approximation. From these potential-energy surfaces the
ground-state $\epsilon_2$ and $\epsilon_4$ deformations are
determined.

\item
The ground-state energy is minimized with respect to $\epsilon_3$ and
also with respect to $\epsilon_6$ for fixed values of  $\epsilon_2$
and  $\epsilon_4$.

\item
When the resulting ground-state shapes have been determined, single-particle
levels are calculated for each nucleus at the appropriate
deformation and stored on disc. The shell-plus-pairing correction is
also calculated and stored on disc at this time. The shell-plus-pairing
correction is then available for use in the calculation of ground-state
masses and in the determination of macroscopic-model constants. It is
the only microscopic quantity required for the mass adjustment.

\item
Now that the ground-state shapes have been determined, the various
shape-dependent functions that occur in the macroscopic energy are
evaluated at each appropriate ground-state shape and stored on disc.

\item
Analogous steps  to those above
 for masses are carried out also for 28
fission-barrier heights.

\item
Least-squares adjustments are now performed,
with the nuclear masses weighted 80\% and the
fission-barrier heights weighted 20\%. The macroscopic-model
constants are determined and the ground-state masses are calculated.

\item
Finally, when the ground-state shapes and masses and
fission-barrier heights are known,
other properties such as $\beta$-decay
half-lives, $\beta$-delayed neutron-emission and fission
probabilities,
and $Q$ values for $\alpha$ decay are calculated.
\end{enumerate}

For the major portion of the potential-energy-surface
calculation we have chosen the following
grid:
\beq
\epsilon_2 =  -0.50\; (0.05)\; 0.50, \; \; \; \; \;
\epsilon_4 =  -0.16\; (0.04)\; 0.16
\eeq{highgrid}
When the ground-state minimum is outside this grid we have used
instead the expanded, but less-dense grid:
\beq
\epsilon_2 =  -1.0\; (0.1)\; 1.0, \; \; \; \; \;
\epsilon_4 =  -0.28\; (0.07)\; 0.28
\eeq{lowgrid}
For large values of $\epsilon_4$ the nuclear shapes develop somewhat
unnatural wiggles. These wiggles can be removed and the energy lowered
by use of higher multipoles in the specification of the nuclear
shape$\,^{30,60})$. We include
in the first step of our calculations one higher
multipole, namely $\epsilon_6$. However, since in this step we want to
consider only two independent shape coordinates, we determine $\epsilon_6$
at each value of $\epsilon_2$ and $\epsilon_4$ by minimizing the macroscopic
potential energy for $^{240}$Pu.
For heavy nuclei the value of $\epsilon_6$ obtained in
such a minimization is approximately independent of the nucleus considered.
On the other hand, for very light nuclei minimization with
respect to $\epsilon_6$ (and in some cases with respect to $\epsilon_4$)
leads to values corresponding to unphysical shapes. These arise because
if the distance across a wiggle on the nuclear surface is of the order of the
range of the Yukawa-plus-exponential folding function, the nuclear
energy increases very little but the Coulomb energy decreases strongly with
increasing deformation. For $\epsilon_6$ we avoid this difficulty
by minimizing the energy for $^{240}$Pu, which is sufficiently large
that also with $\epsilon_6$ distortions included the wiggles on the
surface is larger than the range of the Yukawa-plus-exponential function.
In the light region we avoid unphysical values of $\epsilon_4$ by including
only a physical range of values in our grid.

We use the single-particle states of the folded-Yukawa
single-particle potential to  calculate the shell-plus-pairing corrections
at each grid point. Although the constants of the single-particle potential
depend on $Z$ and $N$\/,
for the determination of the ground-state values of $\epsilon_2$
and $\epsilon_4$
we use the same set of calculated levels
for a region of
neighboring nuclei, since it is too time-consuming to repeat the
diagonalization for each value of $Z$ and $N$\/. However, when the same levels
are used for a moderately large region
of nuclei, the shell correction for a
magic nucleus calculated in this way may differ by 1 MeV
or more from the shell correction calculated with the single-particle
potential appropriate to that particular nucleus.

To overcome this difficulty we proceed by first noting that most
constants of the single-particle potential have been determined for nuclei
close to line of $\beta$-stability. Because of this and because the radius
of the single-particle potential is one of its most important constants, we
reduce the $Z$ and $N$ dependence of the
constants of the microscopic model to an $A$ dependence only.
We next divide the nuclear chart into
regions of suitable size, choosing for each region one set of
single-particle constants. The regions are centered about
the mass numbers $A=16$, 20, 40, 60, 80, 100, 120, 140, 160, 180, 200,
220, 240, 260, 280, 300, 320, and 340. The individual values of $Z$ and $N$
for each region are taken to be the closest integers
corresponding to Green's
approximation$\,^{61})$ to the line of $\beta$-stability, namely
\beq
N - Z = \frac{0.4A^2}{A+200}
\eeq{green}

For each nucleus with a mass number different from one of these
central mass numbers we calculate
the microscopic corrections  for
{\it two} sets of constants. For example, nuclei with $201\leq A \leq 239$
are included in the $A=220$ calculation and nuclei with
$221\leq A \leq 259$ are included in the $A=240$ calculation.
To determine $E_{\rm s+p}$ for, say, a
 nucleus with $A=225$ we linearly interpolate
in terms of $A$ between the result for $A=220$ and the result for
$A=240$. We find that such an interpolation gives results that
agree to within a few-hundred keV with
those obtained with a single-particle potential appropriate
to the specific nuclei concerned.

Once the ground-state values of $\epsilon_2$
and $\epsilon_4$ are determined in this way,
the shell-plus-pairing
corrections are recalculated at these ground-state shapes
with the exact single-particle potential appropriate to
each of the 8979 nuclei. The slight approximation made in calculating
the potential-energy surfaces affects only  the calculation
of the {\it shape} and has
a negligible effect on the final energy.

After the ground-state   $\epsilon_2$ and $\epsilon_4$
deformations are determined,
we investigate
the stability of the ground state with respect to
$\epsilon_3$ and $\epsilon_6$ shape degrees of freedom.
Rather than simultaneously varying  all shape degrees
of freedom, we instead vary $\epsilon_3$ and $\epsilon_6$
separately, with $\epsilon_2$ and
$\epsilon_4$ held fixed at the values previously
determined. When $\epsilon_3$ is varied $\epsilon_6$ is set equal
to the value used in the
original  minimization with respect to $\epsilon_2$ and $\epsilon_4$,
and when $\epsilon_6$  is varied  $\epsilon_3$  is set equal to
zero. The deeper of the two minima obtained in these two minimizations
is selected as the ground state. The
importance of the  $\epsilon_3$ and $\epsilon_6$ shape
degrees of freedom is discussed further in Ref.$\,^{10})$.
Because surface wiggles should not become too small
relative to the range in the Yukawa-plus-exponential function,
the $\epsilon_6$ minimization is
carried out only for nuclei with $A>60$.

After the ground-state shapes are determined, the
shell-plus-pairing corrections and shape-dependent
macroscopic functions are calculated and stored on disc.
The programs that use this information to determine the
macroscopic-model constants and calculate
ground-state masses are then run. Although the least-squares
adjustment is a nonlinear one, it takes only a few minutes to
find the optimum constant set in an 8-constant variation and
to calculate the final mass table. At this point it is relatively
simple to investigate alternative model assumptions.
As an example, we discuss  the results of one such investigation
concerning the effect of varying the microscopic pairing constant
$r_{\rm mic}$.

In our earlier pairing-model studies$\,^{9})$
we determined $r_{\rm mic}$ in Eqs.~(\ref{pairnmic}) and (\ref{pairpmic})
by minimizing the rms deviation between pairing gaps calculated in
the LN model and experimental pairing gaps. An alternative possibility is
to find $r_{\rm mic}$ by minimizing the  error in the {\it mass model}. Because
our
change in the order of the  Strutinsky shell correction
does influence slightly the pairing calculations through the
determination of $G$ from the effective-interaction pairing gap
$\Delta_G$, a small change in $r_{\rm mic}$
could in principle be required to obtain
an optimum pairing calculation. In a study
of how the model error $\sigma_{\rm th}$, the Lipkin-Nogami
pairing gap $\Delta_{\rm LN}$, the theoretical-mass pairing gap
$\Delta_{\rm th.mass}$, and the neutron separation energy $S_{\rm n}$
depend on
$r_{\rm mic}$,
we first calculate the ground-state shell-plus-pairing
corrections for several values of $r_{\rm mic}$. For each of these values
we then determine a set of macroscopic-model constants and
generate a full-fledged mass table. In this process we also obtain
microscopic pairing quantities and neutron separation energies
and compare with experimental values.
Recall that $\Delta_{\rm LN}$ is the sum of the pairing gap $\Delta$ and
the number fluctuation constant $\lambda_2$ that occur in the
Lipkin-Nogami equations. The pairing gap $\Delta_{\rm th.mass}$ is determined
from odd-even {\it theoretical\/} mass differences.
The results are summarized in Table~\ref{tabfit}.
\begin{table}[t]
\begin{small}
\begin{center}
\caption[tabfit]{Determination of the pairing constant \label{tabfit}
$r_{\rm mic}$.\\}
\begin{tabular}{ccccc}
 & & & & \\
\hline\\[-0.12in]
      & $\sigma_{\rm th}$   &  rms              &  rms                  &  rms
  \\
 $r_{\rm mic}$  & Mass model & $\Delta_{\rm LN}$ & $\Delta_{\rm th.mass}$ &
$S_{\rm n}$\\
(MeV) & (MeV)      &  (MeV)            & (MeV)                 &
(MeV)\\[0.08in]
\hline\\[-0.07in]
3.1 & 0.6746 & 0.1740 & 0.2035 &  0.409 \\
3.2 & 0.6694 & 0.1691 & 0.2044 &  0.411 \\
3.3 & 0.6695 & 0.1675 & 0.2091 &  0.416 \\
3.5 & 0.6733 & 0.1745 & 0.2287 &  0.432 \\
\hline
\end{tabular}
\end{center}
\end{small}
\end{table}

Ideally, the minimum deviation should occur for
all quantities at the same value of  $r_{\rm mic}$, which
is almost but not quite the case.
As seen in Table~\ref{tabfit},
all
minima are close to the mass-model minimum at $r_{\rm mic}=3.2$
MeV\@. We therefore
choose this  value of $r_{\rm mic}$ for our microscopic
pairing calculations.
Experimental pairing gaps determined from odd-even mass
differences contain large errors arising from non-smooth
contributions to the mass surface
other than pairing effects, for example, from shape transitions
and gaps in the deformed single-particle level spectra.  Since such
contributions are equally present in $\Delta_{\rm exp}$ and
$\Delta_{\rm th.mass}$, they should cancel out
approximately in the difference between these
two quantities,
if the mass model errors were sufficiently small.
Consequently, the other non-smooth
contributions to the mass surface are not expected to
affect an rms minimization of
$\Delta_{\rm th.mass}$.  It is therefore  of interest to note that our
chosen value of $r_{\rm mic}$ is
intermediate between what would have been
obtained from considering $\Delta_{\rm th.mass}$ and $\Delta_{\rm LN}$
deviations.

The FRDM, which includes Coulomb redistribution effects, is now our
preferred nuclear mass model. Relative to the work described in
Ref.$\,^{10})$,
the following further improvements have been incorporated into
the model.  First, it was found that the $\gamma$ zero-point energy
could not be calculated with sufficient accuracy in our current model.
It is therefore no longer included, whereas the $\epsilon_2$ zero-point
energy is still retained. Second, we have also returned to our original
choice of basis functions corresponding to 12
oscillator shells for all $A$ values, instead of using somewhat fewer
basis functions for lighter nuclei$\,^{10})$.  Third, we now use
an eighth-order Strutinsky shell-correction with range
$\gamma=1.0\hbar\omega_0 B_{\rm s}$ instead of our
earlier choice of a sixth-order
Strutinsky shell correction with the same range coefficient but
no dependence on surface area. The change in
zero-point energy reduced the error in the calculated neutron
separation energies from 0.551 MeV to 0.444 MeV and the error
in the calculated masses from
0.778 MeV$\,^{10})$ to 0.773 MeV\@.  The second and third
improvements further reduced the separation-energy error to 0.411 MeV
and the mass-model error to 0.669 MeV\@.  The rms error for $\Delta_{\rm
th.mass}$ has decreased in a similar manner as the error in $S_{\rm n}$.
Although the effect of the mass-model improvements on $\Delta_{\rm LN}$
is small, the effect on $\Delta_{\rm th.mass}$ is dramatic. Relative to
our earlier pairing calculation$\,^{9})$,
the improvement is more than 20\%. It is no
accident that both $S_{\rm n}$ and $\Delta_{\rm th.mass}$ showed similar
improvements. Both are determined from mass differences between nearby
masses, and such differences dramatically improved when the inaccurate
$\gamma$ zero-point energies were excluded from the calculations.
The constants of the final model were presented
in an earlier section.

As seen in Table \ref{tabfit}, the error in our mass model
is now 0.669 MeV\@.
We have also performed a mass calculation with the FRLDM as the
macroscopic model and identical shell-plus-pairing corrections
as in the FRDM calculation. For the FRLDM the corresponding error is
0.779 MeV\@, which is 16\% higher.

Figure \ref{frdmdev} shows the results of the FRDM calculation.  As
usual, the top part shows the differences between measured masses and
the spherical macroscopic FRDM contributions plotted against the neutron
number $N$, with isotopes of a particular element connected by a
line.  These experimental microscopic corrections are to be
compared with the calculated microscopic corrections plotted
in the middle part of the figure. When the macroscopic and microscopic
parts of the mass calculation are combined and subtracted from the
measured masses, the deviations in the bottom part of the figure
remain.  The trends of the error in the heavy region suggest that this
mass model should be quite reliable for nuclei beyond the current end
of the periodic system.
This has been made all the more plausible by
simulations discussed in Sec.~\ref{4.3} on extrapability.
When $\epsilon_3$ and $\epsilon_6$ shape degrees of freedom
are included in the mass calculations, it becomes clear that the FRLDM,
which does not treat Coulomb redistribution effects, is deficient in
the heavy-element region, as is seen in  Fig.~\ref{frldmdev}.  Thus, our
preferred mass model is now the FRDM\@, which includes compressibility
effects and the associated Coulomb redistribution.

\subsection{Compressibility}

We have earlier$\,^{4})$ studied how the discrepancy
between measured masses
and calculated masses depends on the compressibility constant $K$
and on the new exponential term. In this earlier investigation we used the
1984 version of the FRDM$\,^{5})$. We found that the minimum
error occurred for $K=324$ MeV\@. For this value the rms deviation
between calculated and experimental masses was 0.666 MeV\@. For the
conventional value $K=240$ MeV the rms error was only marginally higher,
namely 0.676 MeV\@. Because of the relative insensitivity of the
rms error to the value of the compressibility constant $K$
we retained$\,^{4})$
the historical value $K=240$ MeV, as we also do here.

It is of interest to investigate the sensitivity of the model error to $K$
also in the current version of the model. In Fig.~\ref{comp} the
solid circles connected by a solid line
show the theoretical error  in the mass model as a function of $1/K$.
For each  value of $K$ the constants in the model  are determined by minimizing
the
same weighted sum of barrier and mass errors that is referred to under point
6 in Sec.~\ref{4.1}. However, only the theoretical error
in the mass model itself is
plotted. The arrow at $K=243$ MeV indicates the optimum value of $K$ obtained
when the compressibility coefficient is varied along with the other constants.
Thus, in our current model we
obtain in a least-squares minimization
a compressibility coefficient that is close to the value $K=240$ MeV that
was adopted from other considerations,
but the determination is clearly subject to a large uncertainty.

We  also investigate how the error for nuclei with $N\geq65$ depends on $K$.
This dependence is shown as solid squares
connected by a long-dashed line.
The model constants have the same values as obtained from the
adjustments corresponding to the solid circles. Thus, {\it no new adjustment\/}
is
performed to this limited region of nuclei; we only investigate the
behavior of the error associated with this region.

Finally, we show as open circles connected by a short-dashed line
the result obtained in an adjustment without
the exponential term. Here the minimum of the weighted sum of mass and
barrier errors occurs at $K=451$ MeV\@. The minimum of the function plotted,
which
is the mass error only, occurs at a slightly higher value of $K$.

The relatively low curvature of the solid curve shows that $K$ cannot be
reliably determined from an adjustment to nuclear masses.
The conventional droplet model value $K=240$ MeV is consistent
with the result we obtain in a least-squares adjustment to
masses and fission barrier heights, but from
the adjustment alone one would not be able to rule out that $K$ has some other
value
in the range from somewhat below 200 MeV to about 500 MeV\@.

The long-dashed curve shows that heavy nuclei
disfavor
high values of the compressibility coefficient. This observation has been made
earlier and was taken as evidence for a
Coulomb redistribution effect$\,^{10})$.

The short-dashed curve giving the results without an exponential term
in the mass model
is moderately incompatible with a compressibility near 240~MeV
and completely rules out a significantly lower value.
However, our preferred treatment of the compressibility
is the formulation that includes the exponential term, in
which treatment the restrictions on $K$ are the much
less severe ones given above.

\subsection{Extrapability \label{4.3}}

One test of the reliability of a nuclear mass model is to compare
deviations between measured and calculated masses in new regions of
nuclei that were not considered when the constants of the model were
determined to deviations in the original region.  This type of analysis
was used earlier by Haustein$\,^{62})$. However, we here
considerably modify his approach. In addition to examining the raw
differences between measured and calculated masses, we use these
differences to determine the {\it model\/} mean discrepancy  $\mu_{\rm
th}$ from the true masses and the {\it model\/} standard deviation
$\sigma_{\rm th}$ around this mean, for new regions of nuclei. Whereas
the raw differences do not show the true behavior of the theoretical
error because errors in the measurements contribute to these
differences, by use of the ideas developed in Sec.~\ref{2p1} we are
able to estimate the {\it true\/} mean $\mu_{\rm th}$ and standard
deviation $\sigma_{\rm th}$ of the theoretical error term $e_{\rm
th}$.

Since our new mass model was developed only recently, we cannot test
its reliability in new regions of nuclei because sufficiently many new
data points are not available. Therefore, we have resorted to a simple
simulation, in which we adjusted the constants in the model  to the
same experimental data set that was  used in our 1981 mass
calculation$\,^{1,2})$.  Consequently, this
calculation is not completely identical to the one on which
Fig.~\ref{frdmdev} is based.  The differences  between  the 351 new
masses that are now measured$\,^{45})$ and the calculated masses
are plotted {\it versus\/} the number of neutrons from
$\beta$-stability in  Fig.~\ref{erfrdm}.  We observe no systematic
increase in the error with increasing neutrons  from
$\beta$-stability.  For the new region of nuclei the square root of the
second central moment is 0.686 MeV\@, compared to 0.671 MeV in the
region where the parameters were adjusted, representing an increase of
only 2\%. In contrast, mass models based on postulated shell-correction
terms  and on a correspondingly larger number of constants normally
diverge outside the region where the constants were
determined$\,^{11,12})$.

To study more quantitatively how the error depends upon the distance
from $\beta$-stability, we introduce bins in the error plots
sufficiently wide to contain about 10--20 points and calculate the mean
error and standard deviation about the mean for each of these bins by
use of the methods described in Sec.~\ref{2p1}.  The results for our
1981 finite-range liquid-drop model and for our 1992 finite-range
droplet model, but adjusted only to the same data set as was used in
our 1981 calculation, are shown in  Fig.~\ref{errsum3}.  For each model
the central, {\it light\/}-gray band representing the original error
region extends one standard deviation on each side of zero.  The solid
dots connected by a thick black line represent the mean of the error
for nuclei that were not considered when the constants in the model
were determined. The {\it dark\/}-gray area extends one standard
deviation on each side of this line.
\begin{table}[t]
\begin{small}
\begin{center}
\caption[taberr]{\baselineskip=12pt\small  Comparison
of errors of two different mass calculations.  The errors
\label{taberr} are tabulated both for the region in which the constants
were originally adjusted  and for a set of new nuclei that were not
taken into account in the determination of the constants of the mass
models. The error ratio is the ratio between the numbers in columns 8
and 3.\\}
\begin{tabular}{lccccccccc}
\hline\\[-0.07in]
& \multicolumn{2}{c}{Original nuclei} &
&\multicolumn{5}{c}{New nuclei} &\\[0.08in]
\cline{2-3} \cline{5-10}\\[-0.07in]
Model & {rms} & $\sigma_{\rm th}$ &
& $N_{\rm nuc}$ &  rms & {$\mu_{\rm th}$} & $\sigma_{\rm th}$ & $\sigma_{\rm
th;\mu = 0}$  &
Error \\
  &         {(MeV)} & (MeV) &
 &  {(MeV)} &  {(MeV)}
&  {(MeV)} &  {(MeV)} & (MeV)
&  {ratio} \\[0.08in]
\hline\\[-0.07in]
FRLDM (1981)    & 0.835&0.831& & 351 & 0.911 & $ -0.321$& 0.826 &0.884    &
\phantom{0}1.06\\
FRDM (1992)  & 0.673& 0.671 & & 351 & 0.735 & $ -0.004$ & 0.686 & 0.686 &
\phantom{0}1.02\\[0.08in]
\hline
\end{tabular}\\[3ex]
\end{center}
\end{small}
\end{table}
The properties of the two models displayed in  Figs.~\ref{errsum3}
are summarized in Table \ref{taberr}.

To test the reliability of the FRDM for extrapolation beyond the
heaviest known elements we have performed a rather severe test in which
we adjust the constants in the model only to data in the region $Z,N
\ge 28$ and $A \le 208$.  There are 1110 known masses in this region
compared to 1654 in the region $Z,N \ge 8$ used in our standard
adjustment. Thus, about one third of all known masses are excluded,
with nuclei removed from both ends of the region of adjustment.  We
then apply the model with these constants to the calculation of all
known masses in our standard region and compare the results to our
standard model in Fig.~\ref{extrasup}.  The error for the known nuclei
is now 0.745 MeV\@, compared to 0.669 MeV with our standard model
adjusted to all known nuclei. Although there is a noticeable increase
of the error in the regions that were not included in the adjustment,
an inspection of Fig.~\ref{extrasup} indicates that the increased error
in the heavy region is not due to a systematic divergence of the mean
error, but rather to a somewhat larger scatter in the error.

In our standard model the mass excesses of  $^{272}$110 and
$^{288}$110  are 133.82~MeV and 165.68~MeV\@, respectively. In our
restricted adjustment we obtain 133.65~MeV and 166.79~MeV\@,
respectively.  Thus, although $^{288}$110 is 80 units in $A$ away from
the last nucleus included in the restricted adjustment, the mass
obtained in this numerical experiment is only about 1 MeV different
from that obtained in the calculation whose constants were adjusted to
nuclei up to 50 units in $A$ closer to the superheavy region.  Since
our standard calculation is adjusted so much closer to the superheavy
region than is the numerical experiment, we feel that it should be
accurate to about an MeV in the superheavy region.  Since models with
and without Coulomb redistribution energies often differ by
considerably more, the masses of superheavy elements could provide very
strong further confirmation of the existence of Coulomb redistribution
effects.  A suitable nucleus for such a test is $^{272}$110. The FRDM,
which includes Coulomb-redistribution effects, predicts a mass excess
of 133.82~MeV for this nucleus, whereas the FRLDM, which does not
include Coulomb-redistribution effects, predicts 136.61~MeV\@.

Figure~\ref{frdmdev} shows that as the lighter region is approached the
error  gradually increases in a systematic way. We have explored this
possibility by first determining the model error for limited regions of
nuclei by use of Eq.~(\ref{name16}). We select $A=25(25)250$ as
centerpoints of the regions and define each region to extend from
$A_{\rm center} -24$ to $A_{\rm center} + 25$. The errors in these
restricted regions are shown as solid circles in  Fig.~\ref{errfua}.
Since the trend of the error looks approximately like $c/A^{\alpha}$ we
have determined the parameters of this assumed error function by use of
the maximum-likelihood equations~(\ref{deveq21}) and (\ref{deveq22}).
We find $c=8.62$ MeV and $\alpha = 0.57$.  The error function
corresponding to these parameters is plotted as a solid line.

\subsection{Fission barriers}
Calculated heights of the outer peak in the fission barrier are compared
to measured values in Table~\ref{tabfis}. The results are also shown
graphically in Fig.~\ref{fisbar}.
\begin{table}[t]
\caption[tabfis]{\baselineskip=12pt\small  Comparison
of experimental and calculated \label{tabfis} fission-barrier heights for
28 nuclei.\\}
\begin{center}
\begin{tabular}{rrrrrr}
\hline\\[-0.07in]
$Z$  & $N$ & $A$ &  Experimental & Calculated & Discrepancy  \\
     &     &     & barrier       & barrier    &              \\
     &     &     & (MeV)         & (MeV)      & (MeV)        \\[0.08in]
\hline\\[-0.07in]
   48 &   61 &  109 &     34.00 &     35.69 &   $       -1.69 $ \\
   66 &   94 &  160 &     27.40 &     27.88 &   $       -0.48 $ \\
   76 &  110 &  186 &     23.40 &     21.21 &      2.19 \\
      &  112 &  188 &     24.20 &     21.07 &      3.13 \\
   80 &  118 &  198 &     20.40 &     19.16 &      1.24 \\
   84 &  126 &  210 &     20.95 &     21.81 &   $       -0.86 $ \\
      &  128 &  212 &     19.50 &     19.69 &   $       -0.19 $ \\
   88 &  140 &  228 &      8.10 &      8.41 &   $       -0.31 $ \\
   90 &  138 &  228 &      6.50 &      7.43 &   $       -0.93 $ \\
      &  140 &  230 &      7.00 &      7.57 &   $       -0.57 $ \\
      &  142 &  232 &      6.30 &      7.63 &   $       -1.33 $ \\
      &  144 &  234 &      6.65 &      7.44 &   $       -0.79 $ \\
   92 &  140 &  232 &      5.40 &      6.61 &   $       -1.21 $ \\
      &  142 &  234 &      5.80 &      6.79 &   $       -0.99 $ \\
      &  144 &  236 &      5.75 &      6.65 &   $       -0.90 $ \\
      &  146 &  238 &      5.90 &      4.89 &      1.01 \\
      &  148 &  240 &      5.80 &      5.59 &      0.21 \\
   94 &  144 &  238 &      5.30 &      4.85 &      0.45 \\
      &  146 &  240 &      5.50 &      4.74 &      0.76 \\
      &  148 &  242 &      5.50 &      5.25 &      0.25 \\
      &  150 &  244 &      5.30 &      5.78 &   $       -0.48 $ \\
      &  152 &  246 &      5.30 &      6.27 &   $       -0.97 $ \\
   96 &  146 &  242 &      5.00 &      4.24 &      0.76 \\
      &  148 &  244 &      5.00 &      5.05 &   $       -0.05 $ \\
      &  150 &  246 &      4.70 &      5.69 &   $       -0.99 $ \\
      &  152 &  248 &      5.00 &      6.07 &   $       -1.07 $ \\
      &  154 &  250 &      4.40 &      5.51 &   $       -1.11 $ \\
   98 &  154 &  252 &      4.80 &      5.31 &   $       -0.51 $ \\
\hline
\end{tabular}
\end{center}
\end{table}
Extensive fission studies based on earlier and current versions of the models
discussed here are presented in
Refs.$\,^{30,31,33,63-68})$.

\subsection{Ground-state masses and deformations}

In the Table  we tabulate our calculated ground-state
deformations in the $\epsilon$ parameterization, the corresponding
coefficients $\beta$ in a spherical-harmonics expansion, the atomic mass
excesses
and microscopic energies calculated  in both the FRDM and FRLDM,
and experimental masses and associated errors that were used in the
adjustment of model constants.

To give an overview, most of our calculated quantities  are plotted
versus $N$ and $Z$ in the form of color contour diagrams.
The calculated ground-state deformations
$\epsilon_2$, $\epsilon_3$, $\epsilon_4$, and $\epsilon_6$ are shown in
 Figs.~\ref{feps2}--\ref{feps6}, and the corresponding
coefficients $\beta_2$,  $\beta_3$, $\beta_4$, and $\beta_6$
are shown in  Figs.~\ref{fbeta2}--\ref{fbeta6}.
We observe some features that are by now  well-known.
For example, the quadrupole deformation parameter $\epsilon_2$ increases by
about
0.05 for each deformed region below the actinide region. Oblate deformations
occur in transition regions on the heavy side of most  deformed regions.
The hexadecapole deformation $\epsilon_4$ is large and negative in
the beginning of deformed regions and large and positive in the
end of deformed regions. The coefficients
  $\beta_3$,  $\beta_4$, and $\beta_6$  have the opposite sign from
the corresponding $\epsilon$ deformations, whereas $\beta_2$ has
the same sign as $\epsilon_2$ but is roughly 10\%  larger.

The microscopic correction is plotted in Fig.~\ref{femic}. The
familiar doubly magic regions around
$^{100}_{\phantom{0}50}$Sn$_{50}$,
$^{132}_{\phantom{0}50}$Sn$_{82}$, and
$^{208}_{\phantom{0}82}$Pb$_{126}$ stand out
clearly. The center of the superheavy region is
located at $^{294}$115$_{179}$. The
large negative microscopic correction originating in the
superheavy
region extends a significant distance towards the southwest and reaches
into the deformed actinide region. It is these large, negative
microscopic corrections that have made possible the extension of
the known elements  as far as $^{266}_{109}$Mt$_{157}$.
As is seen in  Figs.~\ref{fdeps3} and \ref{fdeps6},
the largest effects of $\epsilon_3$ and $\epsilon_6$ in
experimentally accessible regions occur around $^{222}$Ra and
$^{252}$Fm, respectively.

In Fig.~\ref{ferr} we show the discrepancy between
experimental and calculated masses in the form of a contour diagram
versus $N$ and $Z$\@. Above $N\approx65$ there are only a few nuclei with
an error marginally larger than 1 MeV\@. The noticeable errors near
$Z=40$, $N= 56$  are probably related to the unique$\,^{14})$
shell structure in this region
and the reinforcement of the $N=56$ shell closure for proton number $Z=40$
and proton numbers just below.
Such proton-neutron interactions are not accurately described within
any simple single-particle effective-interaction framework.

\section{Acknowledgements}

We are grateful to G.\ Audi for permission to use the results of
his unpublished 1989 midstream atomic mass evaluation.
This work was supported by the U.\ S.\ Department of Energy.  One of us
(P.\ M.) would like to acknowledge through a historical note the
hospitality and support received  during the course of this work.  The
main sponsor of the mass model work during the years 1985--1993 has
been the Los Alamos National Laboratory, but numerous other
institutions have also been involved. During visits to Lawrence
Berkeley Laboratory in the summers of 1981 and 1982 and in the 1983--84
academic year the finite-range droplet model was
developed$\,^{5})$.  Systematic work on an improved mass model
started during visits to Los Alamos National Laboratory in 1985--87,
and interim results were published a year
later$\,^{3,4})$.  The calculation of contour maps for
8979 nuclei was sponsored by Lawrence Livermore National Laboratory in
the fall of 1987. The Lipkin-Nogami pairing code was developed as part
of a contract with Idaho National Engineering Laboratory in 1988.  In
1990 a completely new code for the FRDM was written during a summer
visit to Lund University. Whereas the previous code would run only on
Cray and CDC computers the new code could run on any workstation.
Thus, we were able to carry out the minimization of the potential
energy with respect to $\epsilon_3$ and $\epsilon_6$ without
substantial charges on available workstation clusters.  Initial
minimization calculations were carried out during the visit to Lund
University.  These were continued in the fall of 1990 during a visit to
Institut f\"{u}r Kernchemie, Mainz, which visit was also sponsored by
Gesellschaft f\"{u}r Schwerionenforschung, Darmstadt.  The model
development and calculations were brought to their current stage at Los
Alamos National Laboratory in
1991$\,^{9,10,12})$ and in
1992$\,^{11})$.  This publication was put together at Los Alamos
in the summer of 1993.
\newpage
\markboth
{\it P. M\"{o}ller, J. R. Nix, W. D. Myers, and W. J. Swiatecki/Nuclear Masses}
{\it P. M\"{o}ller, J. R. Nix, W. D. Myers, and W. J. Swiatecki/Nuclear Masses}
\begin{center}
{\bf References}
\end{center}
\newcounter{bona}
\begin{list}%
{\arabic{bona})}{\usecounter{bona}
\setlength{\leftmargin}{0.5in}
\setlength{\rightmargin}{0.0in}
\setlength{\labelwidth}{0.3in}
\setlength{\labelsep}{0.15in}
}
\item
P.\ {M\"{o}ller} and J.\ R.\ Nix, Nucl.\ Phys.\ {\bf A361} (1981) 117.

\item
P.\ {M\"{o}ller} and J.\ R.\ Nix, {Atomic Data Nucl.\ Data Tables} {\bf 26}
  (1981) 165.

\item
P.\ M{\"{o}}ller and J.\ R.\ Nix, {Atomic Data Nucl.\ Data Tables} {\bf 39}
  (1988) 213.

\item
P.\ M{\"{o}}ller, W.\ D.\ Myers, W.\ J.\ Swiatecki, and J.\ Treiner, {Atomic
  Data Nucl.\ Data Tables} {\bf 39} (1988) 225.

\item
P.\ {M\"{o}ller}, W.\ D.\ Myers, W.\ J.\ Swiatecki, and J. Treiner, Proc.\ 7th
  Int.\ Conf.\ on nuclear masses and fundamental constants, Darmstadt-Seeheim,
  1984 (Lehrdruckerei, Darmstadt, 1984) p.\ 457.

\item
W.\ D.\ Myers and W.\ J.\ Swiatecki, Ann.\ Phys.\ (N.~Y.) {\bf 55} (1969) 395.

\item
W.\ D.\ Myers and W.\ J.\ Swiatecki, Ann.\ Phys.\ (N.~Y.) {\bf 84} (1974) 186.

\item
W.\ D.\ Myers, Droplet model of atomic nuclei (IFI/Plenum, New York, 1977).

\item
P.\ M{\"{o}}ller and J.\ R.\ Nix, Nucl.\ Phys.\ {\bf A536} (1992) 20.

\item
P.\ M{\"{o}}ller, J.\ R.\ Nix, W.\ D.\ Myers, and W.\ J.\ Swiatecki, Nucl.\
  Phys.\ {\bf A536} (1992) 61.

\item
P.\ {M\"{o}ller}, J.\ R.\ Nix, K.-L.\ Kratz, A.\ W{\"{o}}hr, and F.-K.\
  Thielemann, Proc.\ 1st Symp.\ on nuclear physics in the universe, Oak Ridge,
  1992 (IOP Publishing, Bristol, 1993) to be published.

\item
P.\ {M\"{o}ller} and J.\ R.\ Nix, Proc.\ 6th Int.\ Conf.\ on nuclei far from
  stability and 9th Int.\ Conf.\ on nuclear masses and fundamental constants,
  Bernkastel-Kues, 1992 (IOP Publishing, Bristol, 1993) p.\ 43.

\item
G.\ A.\ Leander and P.\ M{\"{o}}ller, Phys.\ Lett.\ {\bf 110B} (1982) 17.

\item
R.\ Bengtsson, P.\ {M\"{o}ller}, J.\ R.\ Nix, and Jing-ye Zhang, Phys.\ Scr.\
  {\bf 29} (1984) 402.

\item
J.\ H.\ Hamilton, A.\ V.\ Ramayya, C.\ F.\ Maguire, R.\ B.\ Piercy, R.\
  Bengtsson, P.\ M{\"{o}}ller, J.\ R.\ Nix, Jing-ye Zhang, R.\ L.\ Robinson,
  and S.\ Frauendorf, J.\ Phys.\ G: Nucl.\ Phys.\ {\bf 10} (1984) L87.

\item
P.\ M{\"{o}}ller, G.\ A.\ Leander, and J.\ R.\ Nix, Z.\ Phys.\ {\bf A323}
  (1986) 41.

\item
G.\ A.\ Leander, R.\ K.\ Sheline, P.\ M{\"{o}}ller, P.\ Olanders, I.\
  Ragnarsson, and A.\ J.\ Sierk, Nucl.\ Phys.\ {\bf A388} (1982) 452.

\item
W.\ Nazarewicz, P.\ Olanders, I.\ Ragnarsson, J.\ Dudek, G.\ A.\ Leander, P.\
  M{\"{o}ller}, and E.\ Ruchowska, Nucl.\ Phys.\ {\bf A429} (1984) 269.

\item
G.\ A.\ Leander and Y.\ S.\ Chen, Phys.\ Rev.\ {\bf C37} (1988) 2744.

\item
P.\ M{\"{o}}ller and J.\ R.\ Nix, {Atomic Data Nucl.\ Data Tables} (1993) to be
  published.

\item
H.\ J.\ Lipkin, Ann.\ Phys.\ (N.\ Y.) {\bf 9} (1960) 272.

\item
Y.\ Nogami, Phys.\ Rev.\ {\bf 134} (1964) B313.

\item
H.\ C.\ Pradhan, Y.\ Nogami, and J.\ Law, Nucl.\ Phys.\ {\bf A201} (1973) 357.

\item
D.\ G.\ Madland and J.\ R.\ Nix, Nucl.\ Phys.\ {\bf A476} (1988) 1.

\item
V.\ M.\ Strutinsky, Nucl.\ Phys.\ {\bf A95} (1967) 420.

\item
V.\ M.\ Strutinsky, Nucl.\ Phys.\ {\bf A122} (1968) 1.

\item
W.\ D.\ Myers and W.\ J.\ Swiatecki, Nucl.\ Phys.\ {\bf 81} (1966) 1.

\item
W.\ D.\ Myers and W.\ J.\ Swiatecki, Ark.\ Fys.\ {\bf 36} (1967) 343.

\item
M.\ Bolsterli, E.\ O.\ Fiset, J.\ R.\ Nix, and J.\ L.\ Norton, Phys.\ Rev.\
  {\bf C5} (1972) 1050.

\item
P.\ {M\"{o}ller} and J.\ R.\ Nix, Nucl.\ Phys.\ {\bf A229} (1974) 269.

\item
H.\ J.\ Krappe, J.\ R.\ Nix, and A.\ J.\ Sierk, Phys.\ Rev.\ {\bf C20} (1979)
  992.

\item
J.\ R.\ Nix, Nucl.\ Phys.\ {\bf A130} (1969) 241.

\item
P.\ {M\"{o}ller} and J.\ R.\ Nix, Proc.\ Third IAEA Symp.\ on the physics and
  chemistry of fission, Rochester, 1973, vol.\ I (IAEA, Vienna, 1974) p.\ 103.

\item
P.\ {M\"{o}ller}, S.\ G.\ Nilsson, and J.\ R.\ Nix, Nucl.\ Phys.\ {\bf A229}
  (1974) 292.

\item
S.\ G.\ Nilsson, Kgl.\ Danske Videnskab.\ Selskab.\ Mat.-Fys.\ Medd.\ {\bf
  29}:No.\ 16 (1955).

\item
S.\ E.\ Larsson, S.\ G.\ Nilsson, and I.\ Ragnarsson, Phys.\ Lett.\ {\bf 38B}
  (1972) 269.

\item
S.\ E.\ Larsson, Phys.\ Scripta {\bf 8} (1973) 17.

\item
T.\ Bengtsson and I.\ Ragnarsson, Nucl.\ Phys.\ {\bf A436} (1985) 14.

\item
H.\ J.\ Krappe and J.\ R.\ Nix, Proc.\ Third IAEA Symp.\ on the physics and
  chemistry of fission, Rochester, 1973, vol.\ I (IAEA, Vienna, 1974) p.\ 159.

\item
K.\ T.\ R.\ Davies, A.\ J.\ Sierk, and J.\ R.\ Nix, Phys.\ Rev.\ {\bf C13}
  (1976) 2385.

\item
J.\ Treiner, W.\ D.\ Myers, W.\ J.\ Swiatecki, and M.\ S.\ Weiss, Nucl.\ Phys.\
  {\bf A452} (1986) 93.

\item
D.\ G.\ Madland and J.\ R.\ Nix, Bull.\ Am.\ Phys.\ Soc. {\bf 31} (1986) 799.

\item
E.\ R.\ Cohen and B.\ N.\ Taylor, CODATA Bull.\ No.\ {\bf 63} (1986).

\item
E.\ R.\ Cohen and B.\ N.\ Taylor, Rev.\ Mod.\ Phys.\ {\bf 59} (1987) 1121.

\item
G.\ Audi, Midstream atomic mass evaluation, private communication (1989), with
  four revisions.

\item
D.\ J.\ Vieira, private communication (1990).

\item
X.\ G.\ Zhou, X.\ L.\ Tu, J.\ M.\ Wouters, D.\ J.\ Vieira, K.\ E.\ G.\
  L{\"{o}}bner, H.\ L.\ Seifert, Z.\ Y.\ Zhou, and G.\ W.\ Butler, Phys.\
  Lett.\ {\bf B260} (1991) 285.

\item
N.\ A.\ Orr, W.\ Mittig, L.\ K.\ Fifield, M.\ Lewitowicz, E.\ Plagnol, Y.\
  Schutz, Z.\ W.\ Long, L.\ Bianchi, A.\ Gillibert, A.\ V.\ Belozyorov, S.\ M.\
  Lukyanov, Yu.\ E.\ Penionzhkevich, A.\ C.\ C.\ Villari, A.\ Cunsolo, A.\
  Foti, G.\ Audi, C.\ Stephan, and L.\ Tassan-Got, Phys.\ Lett.\ {\bf B258}
  (1991) 29.

\item
A.\ H.\ Wapstra, Proc.\ 6th Int.\ Conf.\ on nuclei far from stability and 9th
  Int.\ Conf.\ on nuclear masses and fundamental constants, Bernkastel-Kues,
  1992 (IOP Publishing, Bristol, 1993) p.\ 979.

\item
W.\ D.\ Myers, Nucl.\ Phys.\ {\bf 145} (1970) 387.

\item
{\AA}.\ Bohr, B.\ R.\ Mottelson, and D.\ Pines, Phys.\ Rev.\ {\bf 110} (1958)
  936.

\item
S.\ T.\ Belyaev, Kgl.\ Danske Videnskab.\ Selskab.\ Mat.-Fys.\ Medd.\ {\bf
  31}:No.\ 11 (1959).

\item
S.\ G.\ Nilsson and O.\ Prior, Kgl.\ Danske Videnskab.\ Selskab.\ Mat.-Fys.\
  Medd.\ {\bf 32}:No.\ 16 (1961).

\item
W.\ Ogle, S.\ Wahlborn, R.\ Piepenbring, and S.\ Fredriksson, Rev.\ Mod.\
  Phys.\ {\bf 43} (1971) 424.

\item
S.\ G.\ Nilsson, C.\ F.\ Tsang, A.\ Sobiczewski, Z.\ {Szyma\'{n}ski}, S.\
  Wycech, C.\ Gustafson, I.-L.\ Lamm, P.\ {M\"{o}ller}, and B.\ Nilsson, Nucl.\
  Phys.\ {\bf A131} (1969) 1.

\item
J.\ R.\ Nix, Ann.\ Rev.\ Nucl.\ Sci.\ {\bf 22} (1972) 65.

\item
P.\ E.\ Haustein, Atomic Data Nucl.\ Data Tables {\bf 39} (1988) 185.

\item
L.\ Spanier and S.\ A.\ E.\ Johansson, {Atomic Data Nucl.\ Data Tables} {\bf
  39} (1988) 259.

\item
P.\ {M\"{o}ller}, Proc.\ 4th IAEA Symp.\ on physics and chemistry of fission,
  {J\"{u}lich}, 1979, vol.\ I (IAEA, Vienna, 1980) p.\ 283.

\item
I.\ Ragnarsson, Proc.\ Int.\ Symp.\ on future directions in studies of nuclei
  far from stability, Nashville, 1979 (North-Holland, Amsterdam, 1980) p.\ 367.

\item
A.\ E.\ S.\ Green, Nuclear physics (McGraw-Hill, New York, 1955) pp.\ 185, 250.

\item
P.\ E.\ Haustein, Proc.\ 7th Int.\ Conf.\ on nuclear masses and fundamental
  constants, Darmstadt-Seeheim, 1984 (Lehrdruckerei, Darmstadt, 1984) p.\ 413.

\item
P.\ {M\"{o}ller} and J.\ R.\ Nix, Nucl.\ Phys.\ {\bf A272} (1976) 502.

\item
P.\ {M\"{o}ller} and J.\ R.\ Nix, Nucl.\ Phys.\ {\bf A281} (1977) 354.

\item
P.\ M{\"{o}}ller, J.\ R.\ Nix, and W.\ J.\ Swiatecki, Nucl.\ Phys.\ {\bf A469}
  (1987) 1.

\item
P.\ M{\"{o}}ller, J.\ R.\ Nix, and W.\ J.\ Swiatecki, Nucl.\ Phys.\ {\bf A492}
  (1989) 349.

\item
P.\ M{\"{o}l}ler, J.\ R.\ Nix and W.\ J.\ Swiatecki, Proc.\ 50 Years with
  Nuclear Fission, Gaithersburg, 1989 (American Nuclear Society, La Grange
  Park, 1989) p.\ 153.

\item
P.\ M{\"{o}}ller and J.\ R.\ Nix, J.\ Phys.\ G: Nucl.\ Phys.\ (1993) to be
  published.

\end{list}
\markboth
{\it P. M\"{o}ller, J. R. Nix, W. D. Myers, and W. J. Swiatecki/Nuclear Masses}
{\it P. M\"{o}ller, J. R. Nix, W. D. Myers, and W. J. Swiatecki/Nuclear Masses}
\mbox{ } \\ [2ex]
\newpage
\begin{center}
{\Large {\bf Figure captions}}\\[4ex]
\end{center}
\newcounter{bean}
\begin{list}
{\Roman{bean}}{\usecounter{bean}
\setlength{\leftmargin}{1.0in}
\setlength{\rightmargin}{0.0in}
\setlength{\labelwidth}{0.75in}
\setlength{\labelsep}{0.25in}
}
\item[Fig.\ \ref{frdmdev}\hfill]
Comparison of experimental  and calculated microscopic corrections for
1654 nuclei, for a macroscopic model corresponding to the finite-range
droplet model. The bottom part showing the difference between these two
quantities is equivalent to the difference between measured and
calculated ground-state masses. There are almost no systematic errors
remaining for nuclei above $N=65$, for which region the error is only
0.448  MeV\@. The results shown in this figure represent our new mass
model.

\item[Fig.\ \ref{frldmdev}\hfill]
Analogous to  Fig.~\ref{frdmdev}, but for  the finite-range liquid-drop
model, which contains no Coulomb-redistribution terms. This leads to
the systematic negative errors in the heavy region, which
indicate that the calculated masses are systematically too high.

\item[Fig.\  \ref{comp}\hfill]
Relation between the compressibility coefficient $K$ and the
theoretical error in the mass model.  Calculated values are indicated
by symbols, which are connected by curves to guide the eye. In our
standard FRDM, which is our preferred model, the theoretical error
depends only relatively weakly on the compressibility coefficient in
the range 200 ${\rm MeV}<K<500$  MeV, as is shown by the
solid circles. Without the exponential term a relatively high
compressibility coefficient would be required.  The error in the heavy
region, shown by the solid squares, indicates that especially heavy
nuclei disfavor high values of the compressibility coefficient.

\item[Fig.\  \ref{erfrdm}\hfill]
Calculation to  show FRDM  reliability in new regions of nuclei.  Here
we use a smaller set of measured masses to
determine  the constants of the model
than in the full calculation shown in  Fig.~\ref{frdmdev}.  The errors for
nuclei not included in the adjustment are displayed in this figure. The
error is only 2\% larger in the new region compared to that in the
region where the constants were determined. The two largest
deviations occur for $^{23}$O and $^{24}$O, which probably indicates
that this region of light very neutron-rich nuclei is outside the range
of model applicability. Proton number 8 is the lowest value of $Z$ that
we consider in this model.

\item[Fig.\ \ref{errsum3}\hfill]
Comparison of the error behavior for two models applied to new nuclei
{\it versus\/} the number of neutrons from $\beta$-stability.

\item[Fig.\ \ref{extrasup}\hfill]
Test of extrapability of the FRDM
towards the superheavy region. The top part of the figure shows
the error of the standard FRDM\@. In the lower part the error was
obtained from a mass model whose constants were determined from
adjustments to the restricted set of nuclei with $Z,N \ge 28$ and $A
\le 208$. In the light region of nuclei there is no noticeable
divergence of the results obtained in the restricted adjustment. In the
heavy region there is some increase in the spread of the error, but no
systematic divergence of the mean error. Based on the more detailed
discussion in the text we deduce that our calculated masses for the
superheavy elements are accurate to about an  MeV\@.

\item[Fig.\     \ref{errfua} \hfill]
Error in the mass calculation as a function of mass number $A$. The theoretical
error has been determined for limited regions throughout the
periodic system. The error represented by each solid circle
is based on nuclei in a region
that extends 24 mass units below the circle and
25 mass units above the  circle. The points are
well approximated by the function 8.62 MeV/$A^{0.57}$.

\item[Fig.\ \ref{fisbar}\hfill]
Comparison of experimental and calculated fission-barrier heights
for 28 nuclei. Isotopes are connected by lines.

\item[Fig.\     \ref{feps2} \hfill]
Calculated ground-state
values of $|\epsilon_2|$ for 7969 nuclei with $N<200$.
Oblate shapes are indicated with horizontal black lines. About 14
deformed regions stand out, bordered or partially bordered by blue
lines corresponding to  magic nucleon numbers.  The magnitude of the
deformation in the deformed regions increases by about 0.05 with
successively lighter regions or as one goes from  neutron-rich
to proton-rich regions.  Deformed regions above $N=184$ usually
have very low fission barriers, and should have spontaneous-fission
half-lives that are too short to be detectable.

\item[Fig.\     \ref{feps3} \hfill]
Calculated ground-state values of $\epsilon_3$ for 7969 nuclei with $N<200$.
Most nuclei in the investigated region inside the black line
are stable with respect to mass-asymmetric octupole deformations; only 640
nuclei are unstable with respect to these
deformations. The largest effects of experimental significance
are centered around $^{222}_{\phantom{0}88}$Ra$_{134}$.
Effects beyond $N=184$ are of no experimental significance,
since these nuclei are too short-lived to be observed.

\item[Fig.\     \ref{feps4} \hfill]
Calculated  ground-state values of $\epsilon_4$ for 7969 nuclei with $N<200$.
Characteristically, the values are large and negative in the
beginning of major deformed regions and large and positive
in the end of major deformed regions. In accordance with
this general trend, $\epsilon_4$ is large and positive near
the rock of stability in the vicinity of $^{272}$110 near the
end of the deformed ``actinide'' region.

\item[Fig.\     \ref{feps6} \hfill]
Calculated ground-state values of $\epsilon_6$ for 7969 nuclei with $N<200$.
The behavior of $\epsilon_6$ is less regular than that of the lower,
even multipole distortions.

\item[Fig.\     \ref{fbeta2} \hfill]
Calculated  ground-state values of $|\beta_2|$ for  7969 nuclei with $N<200$,
which have been obtained by use of
the transformation~(\ref{betaconv})
from the
$\epsilon$ deformations.
Oblate shapes are indicated by horizontal black lines.
Comments given in Fig.~\ref{feps2} also apply here.

\item[Fig.\     \ref{fbeta3} \hfill]
Calculated  ground-state values of $|\beta_3|$ for  7969 nuclei with $N<200$,
which have been obtained by use of
the transformation~(\ref{betaconv})
from the
$\epsilon$ deformations.
Comments given in Fig.~\ref{feps3} also apply here.

\item[Fig.\     \ref{fbeta4} \hfill]
Calculated  ground-state values of $\beta_4$ for  7969 nuclei with $N<200$,
which have been obtained by use of
the transformation~(\ref{betaconv})
from the
$\epsilon$ deformations.
Comments given in Fig.~\ref{feps4} also apply here,
but note that the sign of $\beta_4$ is opposite that
of $\epsilon_4$ when significant deformations develop.

\item[Fig.\     \ref{fbeta6} \hfill]
Calculated  ground-state values of $\beta_6$ for  7969 nuclei with $N<200$,
which have been obtained by use of
the transformation~(\ref{betaconv})
from the
$\epsilon$ deformations.
Comments given in Fig.~\ref{feps6} also apply here,
but note that the sign of $\beta_6$ is opposite that
of $\epsilon_6$ when significant deformations develop.

\item[Fig.\     \ref{femic} \hfill]
Calculated ground-state microscopic corrections for 7969 nuclei with $N<200$.
Well-known doubly magic regions at $^{100}_{\phantom{0}50}$Sn$_{50}$,
$^{132}_{\phantom{0}50}$Sn$_{82}$, and
$^{208}_{\phantom{0}82}$Pb$_{126}$ stand out clearly.
The minimum in the superheavy region is offset  somewhat from
$^{298}$114$_{184}$ and is  located instead at $^{294}$115$_{179}$.
An interesting feature, also present in our first mass
calculation$\,^{2,14})$, is the rock of stability
at $^{272}_{109}$Mt$_{163}$.

\item[Fig.\     \ref{fdeps3} \hfill]
Calculated ground-state octupole instability  for 7969 nuclei with $N<200$.
Only 640 nuclei exhibit any instability with respect to this shape degree of
freedom. The largest effect in the experimentally
accessible region is $ -1.41$ MeV for $^{222}_{\phantom{0}89}$Ac$_{133}$.

\item[Fig.\     \ref{fdeps6} \hfill]
Calculated ground-state hexacontatetrapole instability  for 7969 nuclei with
$N<200$.
The instability is relative to the
energy corresponding to the macroscopic equilibrium value
of $\epsilon_6$. The largest effect
is $ -1.29$ MeV for $^{251}_{\phantom{0}99}$Es$_{152}$.

\item[Fig.\     \ref{ferr} \hfill]
Discrepancy between measured and calculated masses. Above
$N=65$ only a few discrepancies are marginally more that
1 MeV\@. There is a gradual increase of the error towards the light region.
The large, fluctuating error near $N=60$ is probably due to
deviations between our simple effective interaction and the
true nuclear force. It is well-known that for $Z\approx40$ there
is a re-enforcement of the $N=56$ sub-shell closure. Such effects
cannot be described within the framework of a single-particle model.

\end{list}
\newpage
\setcounter{page}{67}
\begin{center}
{\bf EXPLANATION OF TABLE}
\end{center}
{\bf Table. Calculated Nuclear Ground-State Masses and Deformations,
Compared to Experimental Masses Where Available}\\
\begin{list}
{\Roman{bean}}{\usecounter{bean}
\setlength{\leftmargin}{1.0in}
\setlength{\rightmargin}{0.0in}
\setlength{\labelwidth}{0.75in}
\setlength{\labelsep}{0.25in}
}
\item[{\boldmath $Z$} \hfill ]
Proton number. The mass table is ordered by
increasing
proton number. The corresponding chemical symbol
of named elements is given in parenthesis.

\item[$N$ \hfill ]
Neutron number

\item[$A$ \hfill ]
Mass Number

\item[$\epsilon_2$ \hfill ]
Calculated ground-state
quadrupole deformation in the Nilsson perturbed-spheroid parameterization

\item[$\epsilon_3$ \hfill ]
Calculated ground-state
octupole deformation in the Nilsson perturbed-spheroid parameterization

\item[$\epsilon_4$ \hfill ]
Calculated ground-state
hexadecapole deformation in the Nilsson perturbed-spheroid parameterization

\item[$\epsilon_6$ \hfill ]
Calculated ground-state
hexacontatetrapole deformation in the Nilsson perturbed-spheroid
parameterization

\item[$\epsilon_6^{\rm sym}$ \hfill ]
Calculated ground-state
hexacontatetrapole deformation in the Nilsson perturbed-spheroid
parameterization

\item[$\beta_2$ \hfill ]
Calculated quadrupole deformation of the nuclear ground-state
expressed in the spherical-harmonics expansion~(\ref{betapar})

\item[$\beta_3$ \hfill ]
Calculated octupole deformation of the nuclear ground-state
expressed in the spherical-harmonics expansion~(\ref{betapar})

\item[$\beta_4$ \hfill ]
Calculated hexadecapole deformation of the nuclear ground-state
expressed in the spherical-harmonics expansion~(\ref{betapar})

\item[$\beta_6$ \hfill ]
Calculated hexacontatetrapole deformation of the nuclear ground-state
expressed in the spherical-harmonics expansion~(\ref{betapar})

\item[$E_{\rm mic}$ \hfill ]
Calculated ground-state microscopic energy, given by the difference
between the calculated ground-state atomic mass excess and the spherical
macroscopic
energy calculated from Eq.~(\ref{macener}),
in our preferred model, the FRDM

\item[$M_{\rm th}$ \hfill ]
Calculated ground-state atomic mass excess,
in our preferred model, the FRDM

\item[$M_{\rm exp}$ \hfill ]
Experimental ground-state atomic mass excess
in the 1989 midstream evaluation of Audi$\,^{45})$,
with 4 revisions

\item[$\sigma_{\rm exp}$ \hfill ]
Experimental error associated with the  ground-state atomic mass excess
in the 1989 midstream evaluation of Audi$\,^{45})$,
with 4 revisions

\item[$E_{\rm mic}^{\rm FL}$ \hfill ]
Calculated ground-state microscopic energy, given by the difference
between the calculated ground-state atomic mass excess and the spherical
macroscopic
energy calculated from Eq.~(\ref{macenera}), in the FRLDM

\item[$M_{\rm th}^{\rm FL}$ \hfill ]
Calculated ground-state atomic mass excess, in the FRLDM

\end{list}
\newpage
\begin{enumerate}
\item
    \label{frdmdev}
\item
    \label{frldmdev}
\item
    \label{comp}
\item
    \label{erfrdm}
\item
    \label{errsum3}
\item
    \label{extrasup}
\item
    \label{errfua}
\item
    \label{fisbar}
\item
    \label{feps2}
\item
    \label{feps3}
\item
    \label{feps4}
\item
    \label{feps6}
\item
    \label{fbeta2}
\item
    \label{fbeta3}
\item
    \label{fbeta4}
\item
    \label{fbeta6}
\item
    \label{femic}
\item
    \label{fdeps3}
\item
    \label{fdeps6}
\item
    \label{ferr}
\end{enumerate}
\end{document}